\documentclass[aps,pre,graphicx]{revtex4}

\usepackage[dvips]{graphicx}
\newcommand{\la}{\left\langle}
\newcommand{\ra}{\right\rangle}
\newcommand{\EPL}{{\it Europhys.~Lett.~}}
\newcommand{\PRL}{{\it Phys.~Rev.~Lett.~}}
\newcommand{\PR}{{\it Phys.~Rev.~}}
\newcommand{\JCP}{{\it J.~Chem.~Phys.~}}
\newcommand{\CPL}{{\it Chem.~Phys.~Lett.~}}
\newcommand{\JPC}{{\it J.~Phys.~Chem.~}}
\newcommand{\JSP}{{\it J.~Stat.~Phys.~}}
\newcommand{\JPCM}{{\it J.~Phys.: Condens.~Matter~}}
\newcommand{\MP}{{\it Mol.~Phys.~}}
\newcommand{\JCIS}{{\it J.~Coll.~Int.~Sci.~}}
\newcommand{\EPJ}{{\it Eur.~Phys.~J.~}}

\begin{document}

 
\title{Phase Separation in Charge-Stabilized Colloidal Suspensions: \\
Influence of Nonlinear Screening}

\author{A. R. Denton}
\email{alan.denton@ndsu.edu}
\affiliation{Department of Physics, North Dakota State University,
Fargo, ND, 58105-5566}

\date{\today}
\begin{abstract}
The phase behavior of charge-stabilized colloidal suspensions is modeled by a 
combination of response theory for electrostatic interparticle interactions 
and variational theory for free energies.  Integrating out degrees of freedom
of the microions (counterions, salt ions), the macroion-microion mixture is 
mapped onto a one-component system governed by effective macroion interactions. 
Linear response of microions to the electrostatic potential of the macroions
results in a screened-Coulomb (Yukawa) effective pair potential and a one-body 
volume energy, while nonlinear response modifies the effective interactions 
[A.~R.~Denton, \PR E {\bf 70}, 031404 (2004)].  The volume energy and effective
pair potential are taken as input to a variational free energy, based on 
thermodynamic perturbation theory.  For both linear and first-order nonlinear 
effective interactions, a coexistence analysis applied to aqueous suspensions 
of highly charged macroions and monovalent microions yields bulk separation of 
macroion-rich and macroion-poor phases below a critical salt concentration, 
in qualitative agreement with predictions of related linearized theories 
[R.~van Roij, M.~Dijkstra, and J.-P.~Hansen, \PR E {\bf 59}, 2010 (1999); 
P.~B.~Warren, \JCP {\bf 112}, 4683 (2000)].  
It is concluded that nonlinear screening can modify phase behavior but does not
necessarily suppress bulk phase separation of deionized suspensions.
\end{abstract}


\maketitle




\section{Introduction}
Mounting evidence from a variety of experiments suggests that colloidal 
suspensions~\cite{hunter,pusey,schmitz} of highly charged macroions and 
monovalent microions (counterions and coions) can separate into macroion-rich 
and -poor bulk phases at low salt concentrations.  Reported observations -- in 
aqueous suspensions at sub-millimolar ionic strengths -- describe liquid-vapor 
coexistence~\cite{tata92}, stable voids~\cite{ise94,ise96,tata97,ise99}, 
contracted crystal lattices~\cite{matsuoka,ise99,groehn00}, and metastable 
crystallites~\cite{grier97}.  Such phenomena suggest an unusual form of 
interparticle cohesion, inconsistent with the long-ranged repulsive 
electrostatic pair interactions that prevail at low ionic 
strengths~\cite{israelachvili}, and in apparent conflict with the classic 
theory of Derjaguin, Landau, Verwey, and Overbeek (DLVO)~\cite{DL,VO}, 
which so successfully describes phase stability with respect to coagulation at 
higher salt concentrations.  Observations of bulk phase separation in 
deionized suspensions are therefore often considered anomalous. 

Reports of anomalous phase behavior in charged colloids have been variously
disputed~\cite{palberg94}, attributed to impurities~\cite{belloni00,royall03}, 
or interpreted as genuine manifestations of like-charge interparticle 
attraction~\cite{tata92,ise94,ise96,tata97,ise99,matsuoka,groehn00,grier97}, 
whether pairwise or many-body in origin~\cite{schmitz96,schmitz99,schmitz03}.  
Although some particle-tracking experiments~\cite{grier97,larsen-grier96,
crocker-grier96,fraden94} appear to exhibit attractive forces between isolated 
pairs of tightly confined macroions, recent studies, based on refined optical 
imaging methods, have found no attraction~\cite{bechinger05}.  
Furthermore, mathematical proofs that Poisson-Boltzmann theory predicts purely 
repulsive pair interactions~\cite{neu99,sader-chan,trizac} relegate any 
possible pair attraction to the influence of counterion correlations, neglected
by the mean-field theory.  It is now widely accepted that correlations
among multivalent counterions can induce attraction between like-charged 
surfaces~\cite{jonsson84,guldbrand84,kjellander84,kjellander88,rouzina96}, 
as well as condensation of DNA and other polyelectrolytes~\cite{rouzina96,
ha-liu,levin-comment99,levin99,stevens99,solis99,shklovskii99,nguyen00,levin01,
gelbart01}.  The key issue motivating the present study is whether relatively 
weakly correlated monovalent counterions can similarly destabilize deionized 
colloidal suspensions.  

Further evidence for effective attractive interactions in charged colloids 
comes from computer simulations.  Monte Carlo simulations~\cite{linse-lobaskin,
hynninen05,allahyarov98,messina00} of the primitive model of asymmetric electrolytes 
-- macroions and microions, in a dielectic continuum, directly interacting via 
repulsive Coulomb pair potentials -- exhibit macroion attraction and 
instabilities toward macroion aggregation at high electrostatic couplings.
Short-ranged attractions have been linked to spatial correlations among 
counterions localized near different macroions~\cite{linse-lobaskin,hynninen05}, 
or to Coulomb depletion~\cite{allahyarov98}, while long-ranged attractions have been 
attributed to overcharging of macroions~\cite{messina00}.  System parameters 
thus far explored correspond to relatively strongly correlated (multivalent) 
counterions and relatively small macroion-to-counterion size and charge 
asymmetries.  Computational advances, however, are rapidly closing the gap 
that currently prevents direct comparison of simulations and experiments.

Many theoretical studies of interparticle interactions and phase behavior in 
charged colloids have been motivated by the puzzling results of experiments 
and simulations.  Among various analytical and computational approaches, 
recently reviewed~\cite{belloni00,hansen-lowen00,likos01,levin02}, are 
integral-equation, Poisson-Boltzmann, density-functional, Debye-H\"uckel, 
and response theories.  In seminal work, van Roij {\it et al.}~\cite{vRH97,
vRDH99,vRE99,vR05}, described the phase behavior of charged colloids within 
an effective one-component model governed by density-dependent effective 
interactions.  Combining a linearized density-functional theory~\cite{graf98} 
for the effective pair and one-body (volume energy) potentials with a 
variational theory for the free energy, these authors predicted 
counterion-driven bulk phase separation in deionized suspensions of highly 
charged macroions below a critical salt concentration.  Subsequently, 
Warren~\cite{warren00} applied an extended Debye-H\"uckel (linearized 
Poisson-Boltzmann) theory and predicted similarly unusual phase separation 
at low salt concentrations.  Statistical mechanical~\cite{chan85,chan01} and 
linear-response~\cite{denton99,denton00,denton04} methods, based on closely 
related linearization approximations, yield similar effective electrostatic 
interactions.

Several recent studies, based on Poisson-Boltzmann cell 
models~\cite{vongrunberg01,deserno02,tamashiro03} and extensions of 
Debye-H\"uckel theory~\cite{levin03}, have suggested that predicted 
instabilities of charged colloids towards phase separation may be mere 
artifacts of linearization.  The main purpose of the present study is
to directly test this suggestion by explicitly calculating the effect of 
nonlinear screening on the phase behavior of charged colloids.  Working within 
the framework of the effective one-component model and response 
theory~\cite{denton99,denton00,denton04}, we input nonlinear corrections to 
the effective pair potential and volume energy into an accurate variational 
free energy and analyze thermodynamic phase behavior.  The central conclusion 
of the paper is that nonlinear effects can modify phase behavior of deionized 
suspensions, but do not necessarily suppress counterion-driven phase separation.

Outlining the remainder of the paper, Sec.~\ref{Model} first defines the model 
colloidal suspension.  Section~\ref{Methods} next reviews the response theory 
for effective interactions and describes a variational perturbation theory 
for the free energy.  Section~\ref{Results} presents and discusses numerical 
results -- most importantly, equilibrium phase diagrams obtained from a 
coexistence analysis.  
Finally, Sec.~\ref{Conclusions} summarizes and concludes.

\section{Model System}\label{Model}
\begin{figure}
\includegraphics[width=\columnwidth]{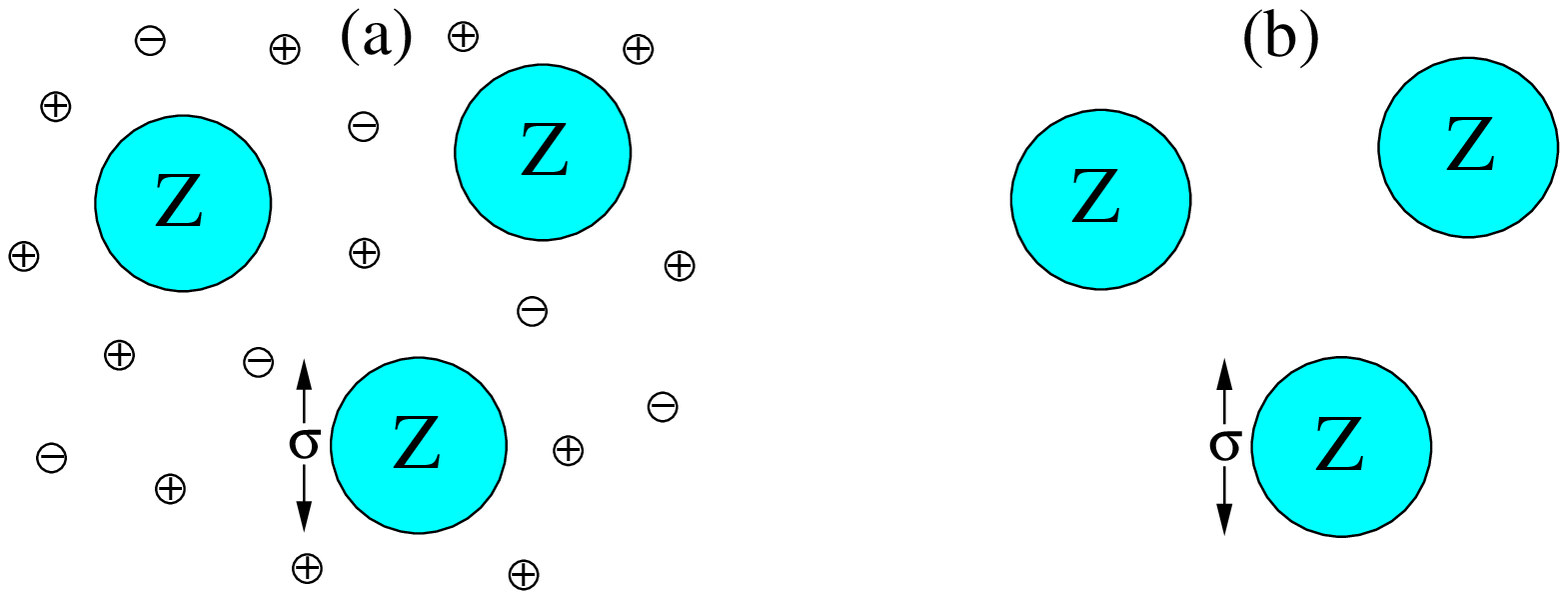}
\vspace*{-15cm}
\caption{\label{fig-model} Models of charge-stabilized colloidal suspensions: 
(a) Primitive model of charged hard-sphere macroions, of effective valence $Z$ 
and diameter $\sigma$, and microions (counterions, salt ions) suspended in 
a dielectric continuum.  (b) Effective one-component model of pseudomacroions 
governed by effective interactions.
}
\end{figure}

The system of interest comprises $N_m$ negatively charged colloidal macroions,
$N_c$ positively charged counterions, and $N_s$ pairs of oppositely charged 
salt ions all dispersed in a solvent.  The macroions are modeled as charged 
hard spheres of radius $a$ (diameter $\sigma$) and effective valence $Z$, as 
depicted in Fig.~\ref{fig-model}.  The macroion surface charge $-Ze$ is best 
interpreted as an effective (renormalized) charge, equal to the bare charge 
less the combined charge of any strongly associated counterions.  The effective
charge is assumed to be uniformly distributed over the surface and fixed, 
independent of thermodynamic state.  The counterions and salt ions are modeled 
as point charges of valence $z$, whose number $N_c$ is determined by the 
condition of overall charge neutrality: $ZN_m=zN_c$.  Numerical results are 
presented below (Sec.~\ref{Results}) for the case of monovalent ($z=1$) 
microions.  The microions number $N_+=N_c+N_s$ positive and $N_-=N_s$ negative,
totaling $N_{\mu}=N_c+2N_s$.  

Working within the primitive model of charged colloids, we approximate 
the solvent as a dielectric continuum, characterized entirely 
by a dielectric constant $\epsilon$.  We further assume a rigid-ion model, 
ignoring van der Waals~\cite{israelachvili} and polarization~\cite{fisher94,
phillies95,gonzalez01} interactions, which are dominated by longer-ranged 
direct electrostatic interactions at low ionic strengths.
The system is imagined to be in thermal equilibrium with a heat bath at 
constant temperature and in chemical (Donnan) equilibrium with a salt reservoir
({\it e.g.}, via a semi-permeable membrane or ion-exchange resin), 
which fixes the salt chemical potential.  
Having specified the model system, we turn next to methods for describing 
electrostatic interactions and thermodynamic phase behavior.

\section{Methods}\label{Methods}
\subsection{Effective Electrostatic Interactions}\label{Interactions}
\subsubsection{One-Component Mapping}\label{Mapping}
Response theory of effective interactions is fundamentally based on mapping 
a multi-component mixture onto a one-component system governed by an effective 
Hamiltonian~\cite{rowlinson84}.  When applied to charged colloids, 
polyelectrolytes, and other ionic systems, the mapping involves integrating out
from the partition function the degrees of freedom of the 
microions~\cite{silbert91}.  The resulting effective interactions between 
macroions depend on the perturbation of the microion distribution by the 
``external" potential of the macroions.  The response of the microions to the 
macroions is linear~\cite{denton99,denton00} for suspensions of 
weakly charged macroions, but becomes increasingly nonlinear~\cite{denton04}  
as the macroion valence increases and as the salt concentration decreases.
Here we briefly review the theory, referring the reader to 
refs.~\cite{denton99,denton00,denton04} for further details.

In the simplest case of a salt-free suspension, the Hamiltonian may be
expressed as
\begin{equation}
H=H_m(\{{\bf R}\})+H_c(\{{\bf r}\})+H_{mc}(\{{\bf R}\},\{{\bf r}\}), 
\label{H}
\end{equation}
where $\{{\bf R}\}$ and $\{{\bf r}\}$ denote coordinates of macroions and 
microions, respectively, 
\begin{equation}
H_m=H_{\rm HS}(\{{\bf R}\})+
\frac{1}{2}\sum_{{i\neq j=1}}^{N_m} v_{mm}(|{\bf R}_i-{\bf R}_j|)
\label{Hm1}
\end{equation}
is the Hamiltonian of the macroions alone, $H_{\rm HS}$ is the Hamiltonian of
a hard-sphere (HS) system, $v_{mm}(r)=Z^2e^2/\epsilon r$, $r>\sigma$, is the 
bare Coulomb pair potential between macroions,
\begin{equation}
H_c=K_c+\frac{1}{2}\sum_{{i\neq j=1}}^{N_c} v_{cc}(|{\bf r}_i-{\bf r}_j|)
\label{Hc1}
\end{equation}
is the counterion Hamiltonian, $K_c$ is the counterion kinetic energy, 
$v_{cc}(r)=z^2e^2/\epsilon r$ is the pair potential between counterions, 
\begin{equation}
H_{mc}=\sum_{i=1}^{N_m}\sum_{j=1}^{N_c} v_{mc}(|{\bf R}_i-{\bf r}_j|)
\label{Hmc1}
\end{equation}
is the total macroion-counterion interaction energy, and
$v_{mc}(r)=Zze^2/\epsilon r$, $r>a$, is the macroion-counterion pair potential.

The mapping from the macroion-counterion mixture to an effective one-component
system of pseudomacroions begins with the canonical partition function 
\begin{equation}
{\cal Z}=\la\la\exp(-\beta H)\ra_c\ra_m, 
\label{part1}
\end{equation}
where $\beta\equiv 1/(k_BT)$ at temperature $T$ and angular brackets 
denote classical traces over counterion ($c$) and macroion ($m$) coordinates. 
The mapping proceeds by formally tracing over the counterion coordinates:
\begin{equation}
{\cal Z}=\la\exp(-\beta H_{\rm eff})\ra_m,
\label{part2}
\end{equation}
where $H_{\rm eff}=H_m+F_c$ is the effective one-component Hamiltonian and 
\begin{equation}
F_c=-k_BT\ln\la\exp\left[-\beta(H_c+H_{mc})\right]\ra_c
\label{Fc1}
\end{equation}
is the free energy of a nonuniform gas of counterions in the presence of 
the fixed macroions.  
Within perturbation theory~\cite{silbert91,HM}, the counterion free energy
can be expressed as
\begin{equation}
F_c=F_0+\int_0^1{\rm d}\lambda\,\frac{\partial F_c(\lambda)}{\partial\lambda}
=F_0+\int_0^1{\rm d}\lambda\,\la H_{mc}\ra_{\lambda},
\label{Fc2}
\end{equation}
where $F_c(\lambda)\equiv -k_BT\ln\la\exp\left[-\beta(H_c+\lambda H_{mc})
\right]\ra_c$, $F_0\equiv F_c(0)=-k_BT\ln\la\exp(-\beta H_c)\ra$ is the 
unperturbed counterion free energy in the case of uncharged 
(yet volume-excluding) macroions, $\la~\ra_{\lambda}$ denotes a counterion 
trace with the macroions charged to a fraction $\lambda$ of their full 
charge, and the $\lambda$-integral charges up the macroions.
After formally adding and subtracting the energy of a uniform compensating 
negative background $E_b$, Eq.~(\ref{Fc2}) becomes
\begin{equation}
F_c=F_{\rm OCP}+\int_0^1{\rm d}\lambda\,\la H_{mc}\ra_{\lambda}-E_b,
\label{Fc3}
\end{equation}
where $F_{\rm OCP}=F_0+E_b$ is the free energy of a classical one-component 
plasma (OCP) in the presence of neutral hard spheres.  
The background and counterions alike are excluded from the hard cores of 
the macroions and therefore occupy a free volume $V'=V(1-\eta)$, where
$\eta=(\pi/6)(N_m/V)\sigma^3$ is the macroion volume fraction.

\subsubsection{Response Theory}\label{Response}
To make practical use of the one-component mapping, the counterion free 
energy must be approximated, for which purpose response theory provides 
a powerful framework.  Because it proves more convenient to manipulate
Fourier components of densities and pair potentials, we first note that 
the macroion Hamiltonian [Eq.~(\ref{Hm1})] and macroion-counterion 
interaction [Eq.~(\ref{Hmc1})] can be equivalently expressed as
\begin{equation}
H_m=H_{\rm HS}+\frac{1}{2V'}\sum_{\bf k} \hat v_{mm}(k) 
[\hat\rho_m({\bf k})\hat\rho_m(-{\bf k})-N_m]
\label{Hm2}
\end{equation}
and
\begin{equation}
H_{mc}=\frac{1}{V'}\sum_{\bf k} \hat v_{mc}(k) 
\hat\rho_m({\bf k})\hat\rho_c(-{\bf k}),
\label{Hmc3}
\end{equation}
where the Fourier transform and its inverse are defined as
\begin{eqnarray}
\hat\rho_m({\bf k})=\int{\rm d}{\bf r}\,\rho_m({\bf r})
e^{-i{\bf k}\cdot{\bf r}} \label{FTa}
\\
\rho_m({\bf r})=\frac{1}{V'}\sum_{\bf k}\hat\rho_m({\bf k})
e^{i{\bf k}\cdot{\bf r}}. \label{FTb}
\end{eqnarray}
Equation~(\ref{Hmc3}) makes evident that $H_{mc}$ depends, through 
$\hat\rho_c({\bf k})$, on the response of the counterion density to 
the macroion charge density.  The counterion response can be approximated
by expanding the ensemble-averaged induced counterion density in a 
functional Taylor series in powers of the dimensionless macroion 
potential~\cite{HM,AS,hafner87,ashcroft66}, 
$u({\bf r})=-\beta\int{\rm d}{\bf r}'\,v_{mc}(|{\bf r}-{\bf r}'|)
\rho_m({\bf r}')$.
Expanding about zero macroion charge ($u=0$), the counterion density 
can be expressed, in Fourier space, as~\cite{denton04}
\begin{eqnarray}
\la\hat\rho_c({\bf k})\ra~&=&~\chi(k)
\hat v_{mc}(k)\hat\rho_m({\bf k})+\frac{1}{V'}\sum_{{\bf k}'}
\chi'({\bf k}',{\bf k}-{\bf k}')
\hat v_{mc}(k') \hat v_{mc}(|{\bf k}-{\bf k}'|) \nonumber \\
~&\times&~\hat\rho_m({\bf k}') \hat\rho_m({\bf k}-{\bf k}')
+\cdots, \qquad k\neq 0, \label{rhock}
\end{eqnarray}
where $\chi$ and $\chi'$ are, respectively, the linear and first nonlinear 
response functions of the uniform OCP.  The response functions are directly
related to the structure of the OCP according to $\chi(k)=-\beta n_c S(k)$ 
and $\chi'({\bf k}',{\bf k}-{\bf k}')=
(\beta^2n_c/2)S^{(3)}({\bf k}',{\bf k}-{\bf k}')$, where 
\begin{equation}
S^{(n)}({\bf k}_1,\cdots,{\bf k}_{n-1})=\frac{1}{N_c}\la\hat\rho_c({\bf k}_1)
\cdots\hat\rho_c({\bf k}_{n-1})\hat\rho_c(-{\bf k}_1\dots-{\bf k}_{n-1})\ra
\label{Sn}
\end{equation}
is the OCP $n$-particle static structure factor~\cite{HM}, 
$S(k)\equiv S^{(2)}(k)$, and $n_c=N_c/V'$ is the average density of 
counterions in the free volume.
The first term on the right side of Eq.~(\ref{rhock}), which is linear in
$\hat v_{mc}(k)$ and $\hat\rho_m({\bf k})$, represents the linear response 
approximation, while the higher-order terms are nonlinear corrections.

Combining Eqs.~(\ref{Fc3}), (\ref{Hmc3}), and (\ref{rhock}), specifying the 
background energy as $E_b=\lim_{k\to 0}\{-N_c n_c\hat v_{cc}(k)/2\}$, 
isolating the $k=0$ terms, and integrating over $\lambda$, produces the 
counterion free energy to third order in the macroion density:
\begin{eqnarray}
& &F_c=F_{\rm OCP} + n_c\lim_{k\to 0}
\left[N_m\hat v_{mc}(k) + \frac{N_c}{2}\hat v_{cc}(k)\right]
+ \frac{1}{2V'}\sum_{{\bf k}\neq 0}\chi(k)\left[\hat v_{mc}(k)\right]^2
\hat\rho_m({\bf k})\hat\rho_m(-{\bf k}) \nonumber \\
&+&\frac{1}{3V'^2}\sum_{{\bf k}\neq 0}\sum_{{\bf k}'}\chi'({\bf k}',
-{\bf k}-{\bf k}')\hat v_{mc}(k)\hat v_{mc}(k')\hat v_{mc}(|{\bf k}+{\bf k}'|)
\hat\rho_m({\bf k})\hat\rho_m({\bf k}')\hat\rho_m(-{\bf k}-{\bf k}').
\label{Fc4}
\end{eqnarray}
The terms in $F_c$ that are quadratic and cubic in $\hat\rho_m({\bf k})$ 
generate effective pair and triplet interactions, respectively, 
in the effective Hamiltonian.  To demonstrate this, we first identify
\begin{equation}
\hat v^{(2)}_{\rm ind}(k)=\chi(k)[\hat v_{mc}(k)]^2
\label{v2indk}
\end{equation}
as an effective pair interaction, induced by linear response of 
counterions~\cite{silbert91,denton99,denton00}, and
\begin{equation}
\hat v^{(3)}_{\rm eff}({\bf k},{\bf k}')=2\chi'({\bf
k}',-{\bf k}-{\bf k}') \hat v_{mc}(k)\hat v_{mc}(k')\hat
v_{mc}(|{\bf k}+{\bf k}'|) 
\label{v3effk}
\end{equation}
as an effective three-body interaction, induced by nonlinear counterion 
response.  Combining Eqs.~(\ref{Hm2}) and (\ref{Fc4}), the effective 
Hamiltonian now can be recast in the form
\begin{equation}
H_{\rm eff}=H_{\rm HS}+\frac{1}{2}\sum_{i\neq j=1}^{N_m}
v^{(2)}_{\rm eff}(|{\bf R}_i-{\bf R}_j|)+
\frac{1}{3!}\sum_{i\neq j\neq k=1}^{N_m} 
v^{(3)}_{\rm eff}({\bf R}_i-{\bf R}_j,{\bf R}_i-{\bf R}_k)+E, 
\label{Heff2}
\end{equation}
where $v^{(2)}_{\rm eff}(r)=v_{mm}(r)+v^{(2)}_{\rm ind}(r)$ and 
$v^{(3)}_{\rm eff}({\bf r},{\bf r}')$ 
are the effective macroion pair and triplet potentials, respectively,
and $E$ is a one-body volume energy, composed of all terms in $H_{\rm eff}$ 
independent of macroion coordinates.  The volume energy accounts for the 
counterion entropy and macroion-counterion interaction energy and contributes 
density-dependent terms to the total free energy that can influence 
thermodynamic properties, as discussed below (Sec.~\ref{Results}).

Explicit expressions for the effective interactions are obtained by 
invoking the identities
\begin{equation}
\sum_{{i\neq j=1}}^{N_m} v^{(2)}_{\rm ind}(|{\bf R}_i-{\bf R}_j|)
=\frac{1}{V'}\sum_{{\bf k}\neq 0} \hat v^{(2)}_{\rm ind}(k)
\hat\rho_m({\bf k})\hat\rho_m(-{\bf k})
+\frac{N_m^2}{V'}\lim_{k\to 0}\hat v^{(2)}_{\rm ind}(k)
-N_m v^{(2)}_{\rm ind}(0)
\label{Fc4-3}
\end{equation}
and
\begin{eqnarray}
\sum_{{i\neq j\neq k=1}}^{N_m} v^{(3)}_{\rm eff}
({\bf R}_i-{\bf R}_j,{\bf R}_i-{\bf R}_k)
~&=&~\frac{1}{V'^2}
\sum_{\bf k}\sum_{{\bf k}'}\hat v^{(3)}_{\rm eff}({\bf k},{\bf k}')
[\hat\rho_m({\bf k})\hat\rho_m({\bf k}')\hat\rho_m(-{\bf k}-{\bf k}')
\nonumber \\
~&-&~3\hat\rho_m({\bf k})\hat\rho_m(-{\bf k})+2N_m].
\label{Fc4-4}
\end{eqnarray}
The volume energy, $E=E_0+\Delta E$, is the sum of the linear response 
approximation~\cite{denton99,denton00} 
\begin{equation}
E_0=F_{\rm OCP}+\frac{N_m}{2}v^{(2)}_{\rm ind}(0) 
+N_mn_c\lim_{k\to 0}\left[\hat v_{mc}(k)
-\frac{z}{2Z}\hat v^{(2)}_{\rm ind}(k)+\frac{Z}{2z}\hat v_{cc}(k)\right]
\label{E01}
\end{equation}
and the first nonlinear correction~\cite{denton04}
\begin{equation}
\Delta E=\frac{N_m}{6V'^2}\left[\sum_{{\bf k},{\bf k}'} \hat
v^{(3)}_{\rm eff}({\bf k},{\bf k}')-N_m\sum_{{\bf k}} \hat
v^{(3)}_{\rm eff}({\bf k},0)\right]. 
\label{Delta-E1}
\end{equation}
Similarly, the effective pair interaction,
$v^{(2)}_{\rm eff}(r)=v^{(2)}_0(r)+\Delta v^{(2)}_{\rm eff}(r)$
is the sum of the linear response approximation~\cite{denton99,denton00}, 
$v^{(2)}_0(r)=v_{mm}(r)+v^{(2)}_{\rm ind}(r)$,
and the first nonlinear correction, $\Delta v^{(2)}_{\rm eff}(r)$,
whose Fourier transform is 
\begin{equation}
\Delta\hat v^{(2)}_{\rm eff}(k)=\frac{1}{V'}\sum_{{\bf k}'}\hat
v^{(3)}_{\rm eff}({\bf k},{\bf k}')-\frac{N_m}{3V'}\hat
v^{(3)}_{\rm eff}({\bf k},0).
\label{Delta-v2effk}
\end{equation}
It is important to note that nonlinear counterion response generates not only
effective many-body interactions, but also corrections to the effective pair 
and one-body interactions.  It is these corrections [Eqs.~(\ref{Delta-E1}) and
(\ref{Delta-v2effk})] whose impact on phase behavior we examine below in 
Sec.~\ref{Results}.  Note that the final terms on the right sides of 
Eqs.~(\ref{E01})-(\ref{Delta-v2effk}) originate from the charge neutrality 
condition, which required special treatment of the $k=0$ terms in 
Eqs.~(\ref{Fc4}) and (\ref{Fc4-3}).  A simple physical interpretation of 
microion response and its connection to microion-induced effective interactions
between macroions is discussed in ref.~\cite{denton04}.

\subsubsection{Random Phase Approximation}\label{RPA}
Further progress requires specifying the OCP response functions.  To this end, 
we note first that the counterions are usually characterized by relatively
small electrostatic coupling parameters, $\Gamma=\lambda_B/a_c \ll 1$, 
where $\lambda_B=\beta z^2e^2/\epsilon$ is the Bjerrum length and 
$a_c=(3/4\pi n_c)^{1/3}$ is the counterion-sphere radius.  In such 
weakly-coupled plasmas, short-range correlations are often weak enough to 
justify a random phase approximation (RPA)~\cite{HM}, whereby the 
two-particle direct correlation function (DCF) is approximated by its exact 
asymptotic limit: $c^{(2)}(r)=-\beta v_{cc}(r)$ or
$\hat c^{(2)}(k)=-4\pi\beta z^2 e^2/\epsilon k^2$. 
The OCP linear and first nonlinear response functions then take the simple 
analytical forms
\begin{equation}
\chi(k)=\frac{-\beta n_c}{1+\kappa^2/k^2}
\label{chi1}
\end{equation}
and
\begin{equation}
\chi'({\bf k},{\bf k}')=-\frac{k_BT}{2n_c^2}\chi(k)\chi(k')
\chi(|{\bf k}+{\bf k}'|),
\label{chi2}
\end{equation}
where $\kappa=\sqrt{4\pi n_cz^2e^2/\epsilon k_BT}$ is
the Debye screening constant (inverse screening length).
Higher-order nonlinear response leads to higher-order terms 
in the effective Hamiltonian [Eq.~(\ref{Heff2})], which are here neglected.

Practical expressions for the effective interactions follow from specifying the 
macroion-counterion interaction inside the macroion core so as to minimize
counterion penetration -- a strategy similar to that of the pseudopotential 
theory of simple metals~\cite{hafner87,ashcroft66}.  The choice
\begin{equation}
v_{mc}(r)=-\frac{Zze^2\kappa}{\epsilon(1+\kappa a)}, \qquad r<a,
\label{vmcr}
\end{equation}
ensures zero counterion penetration ($\rho_c(r)=0$, $r<a$) at the level of
linear response~\cite{vRH97,denton99,denton00} and virtually eliminates 
counterion penetration in the case of nonlinear response~\cite{denton04}.  
Substituting the Fourier transform of Eq.~(\ref{vmcr}),
\begin{equation}
\hat v_{mc}(k)=-\frac{4\pi Zze^2}{\epsilon(1+\kappa a)k^2}
\left[\cos(ka)+\frac{\kappa}{k} \sin(ka) \right],
\label{vmck2}
\end{equation}
into Eqs.~(\ref{v2indk}), (\ref{v3effk}), and (\ref{E01})-(\ref{Delta-v2effk}) 
then yields the effective interactions.

Upon reintroducing salt ions as a second species of microion~\cite{denton00},
analytical expressions are obtained~\cite{denton04} for the volume energy
and the effective pair potential.  The volume energy is the sum of
the linear response approximation
\begin{equation}
E_0=F_{\rm plasma}-N_m\frac{Z^2e^2}{2\epsilon}\frac{\kappa}{1+\kappa a}
-\frac{k_BT}{2}\frac{(N_+-N_-)^2}{N_++N_-}
\label{E03}
\end{equation}
and the first nonlinear correction
\begin{equation}
\Delta E=\frac{N_mk_BT}{6}\frac{(n_+-n_-)}{n_{\mu}^3}\left[
\frac{Z^2\kappa^3n_{\mu}}{8\pi}\left(\frac{1}{1+\kappa a}\right)^2
-\frac{Z^3\kappa^6}{(4\pi)^2}
\left(\frac{e^{\kappa a}}{1+\kappa a}\right)^3
{\rm E}_1(3\kappa a)\right],
\label{Delta-E1r-analytical}
\end{equation}
where $F_{\rm plasma}=k_BT[N_{\rm +}\ln(n_{\rm +}\Lambda^3)+N_{\rm -}
\ln(n_{\rm -}\Lambda^3)]$ is the ideal-gas free energy of the plasma, 
$n_{\pm}=N_{\pm}/V'$ and $n_{\mu}=N_{\mu}/V'=n_++n_-=n_c+2n_s$ are the
microion number densities in the free volume, $\Lambda$ is the 
thermal wavelength of the microions, and 
\begin{equation}
\kappa=\left(\frac{4\pi z^2e^2 n_{\mu}}{\epsilon k_BT}\right)^{1/2}
=\left(\frac{4\pi z^2e^2}{\epsilon k_BT}\frac{(N_c+2N_s)}
{V(1-\eta)}\right)^{1/2}
\label{kappa}
\end{equation}
is the Debye screening constant, which depends on the {\it total} density
of microions, adjusted for macroion excluded volume.
The effective pair potential is the sum of
\begin{equation}
v^{(2)}_0(r)=\frac{Z^2e^2}{\epsilon}\left(\frac{e^{\kappa a}}
{1+\kappa a}\right)^2~\frac{e^{-\kappa r}}{r}, \qquad r>\sigma, 
\label{v20r}
\end{equation}
which is identical to the DLVO potential with a density-dependent
screening constant, and
\begin{equation}
\Delta v^{(2)}_{\rm eff}(r)=f_1(r)~\frac{e^{-\kappa r}}{r}
+f_2(r)~\frac{e^{\kappa r}}{r}+f_3(r)~\frac{e^{-\kappa a}}{r}, \qquad r>\sigma,
\label{Delta-v2effr-analytical}
\end{equation}
where
\begin{equation}
f_1(r)=C_1\left[\kappa(r-\sigma)+1-e^{-\kappa\sigma}\right]
+C_2\left[{\rm E}_1\left(\kappa(r-a)\right)
+{\rm E}_1\left(3\kappa a\right)
-{\rm E}_1\left(\kappa a\right)\right],
\end{equation}
\begin{equation}
f_2(r)=-C_2~{\rm E}_1\left(3\kappa(r+a)\right),
\end{equation}
\begin{equation}
f_3(r)=C_2~\left[{\rm E}_1(2\kappa(r+a))-{\rm E}_1(2\kappa(r-a))\right],
\label{f3}
\end{equation}
\begin{equation}
C_1=\frac{1}{6}\frac{(n_+-n_-)}{n_{\mu}}\frac{Z^2e^2}{\epsilon}\left(
\frac{e^{\kappa a}}{1+\kappa a}\right)^2,
\label{C1}
\end{equation}
\begin{equation}
C_2=\frac{1}{8\pi}\frac{(n_+-n_-)}{n_{\mu}^2}\frac{Z^3e^2\kappa^3}{z\epsilon}
\left(\frac{e^{\kappa a}}{1+\kappa a}\right)^3,
\label{C2}
\end{equation}
and 
\begin{equation}
{\rm E}_1(x)=\int_1^{\infty}{\rm d}u\,\frac{e^{-xu}}{u}, \qquad x>0,
\label{E1x}
\end{equation}
is the exponential integral function.
The effective three-body interaction can be computed from the generalizations
of Eqs.~(\ref{v3effk}), (\ref{chi2}), and (\ref{vmck2}), with the 
result~\cite{denton04}
\begin{equation}
v^{(3)}_{\rm eff}({\bf r}_1-{\bf r}_2,{\bf r}_1-{\bf r}_3)
=-k_BT\frac{(n_+-n_-)}{n_{\mu}^3}\int{\rm d}{\bf r}\,
\rho_1(|{\bf r}_1-{\bf r}|)\rho_1(|{\bf r}_2-{\bf r}|)
\rho_1(|{\bf r}_3-{\bf r}|), 
\label{v3effr-salt}
\end{equation}
where 
\begin{equation}
\rho_1(r)=\left\{ \begin{array}
{l@{\quad\quad}l}
\frac{\displaystyle Z}{\displaystyle z}\frac{\displaystyle \kappa^2}
{\displaystyle 4\pi}
~\frac{\displaystyle e^{\kappa a}}{\displaystyle 1+\kappa a}
~\frac{\displaystyle e^{-\kappa r}}{\displaystyle r}, & r>a, \\
0, & r<a,
\end{array} \right.
\label{rho1r}
\end{equation}
is the density of counterions around an isolated macroion.  

It is important to establish the accuracy of the effective interactions
predicted by the nonlinear response theory described above.  In a direct 
comparison with {\it ab initio} simulations~\cite{tehver99}, first-order 
nonlinear corrections were shown to quantitatively match effective pair 
energies~\cite{denton04}.  Nevertheless, the effective interactions predicted 
by response theory should be tested further, perhaps by comparisons with 
nonlinear Poisson-Boltzmann theory.

\subsection{Thermodynamic Phase Behavior}\label{PhaseBehavior}
\subsubsection{Variational Theory}\label{Variational}
The effective electrostatic interactions predicted by response theory provide
the basic input required by statistical mechanical theories and computer 
simulations of the effective one-component model of charged colloids.  
The one-component model is considerably simpler than the (multi-component)
primitive model, and thus a practical alternative for investigating 
thermodynamic phase behavior and other bulk properties of many-particle 
systems.  Here we input effective interparticle interactions into an 
approximate variational theory for the free energy.  The Helmholtz free energy 
$F$ separates naturally into three contributions:
\begin{equation}
F(T,V,N_m,N_s)=F_{\rm id}(T,V,N_m)+F_{\rm ex}(T,V,N_m,N_s)+E(T,V,N_m,N_s),
\label{Ftot}
\end{equation}
where $F_{\rm id}=N_mk_BT[\ln(n_m\Lambda_m^3)-1]$ is the exact 
ideal-gas free energy of a uniform fluid of macroions of thermal wavelength 
$\Lambda_m$, $F_{\rm ex}$ is the excess free energy, which depends on 
effective intermacroion interactions, and $E$ is the one-body volume energy. 
Note that $F_{\rm ex}$ and $E$ depend on the average densities of both 
macroions and salt ions.

To approximate the excess free energy, we apply a variational approach based 
on first-order thermodynamic perturbation theory, as in ref.~\cite{vRDH99}.
Given a decomposition of the effective pair potential into reference 
and perturbation potentials, 
\begin{equation}
v_{\rm eff}^{(2)}(r)=v_{\rm ref}^{(2)}(r)+v_{\rm pert}^{(2)}(r),
\label{decomp}
\end{equation}
an upper bound on the excess free energy density, $f_{\rm ex}=F_{\rm ex}/V$,
is provided by the Gibbs-Bogoliubov inequality~\cite{HM}
\begin{equation}
f_{\rm ex} \leq f_{\rm ref}+\frac{1}{2}n_m^2\int{\rm d}{\bf r}\,g_{\rm ref}(r)
v_{\rm pert}^{(2)}(r),
\label{gb1}
\end{equation}
where $f_{\rm ref}$ and $g_{\rm ref}(r)$ are the excess free energy density 
and radial distribution function, respectively, of the reference system.
The short-range-repulsive form of the effective pair potential naturally 
suggests a hard-sphere (HS) reference system.  Thus,
$v_{\rm ref}^{(2)}(r)=v_{\rm HS}(r;d)$, the pair potential between 
hard spheres of effective diameter $d$, and
$v_{\rm pert}^{(2)}(r)=v_{\rm eff}^{(2)}(r)$, $r\ge d$.
The effective HS diameter provides a variational parameter 
with respect to which the right side of Eq.~(\ref{gb1}) can be minimized 
to impose a least upper bound on the excess free energy:
\begin{equation}
f_{\rm ex}(n_m,n_s) \simeq \min_{(d)}\left\{f_{\rm HS}(n_m,n_s;d)+2\pi n_m^2 
\int_d^{\infty}{\rm d}r\, r^2g_{\rm HS}(r,n_m;d)v_{\rm eff}^{(2)}(r,n_m,n_s)
\right\}.
\label{gb2}
\end{equation}
Here $f_{\rm HS}(n_m,n_s;d)$ and $g_{\rm HS}(r,n_m;d)$ are, respectively, 
the excess free energy density and radial distribution function of the HS 
reference fluid, which we approximate by the essentially exact 
Carnahan-Starling and Verlet-Weis analytical expressions~\cite{HM}.  
In practice, the exponential decay of $v_{\rm eff}^{(2)}(r)$ with $r$ 
ensures rapid convergence of the perturbation integral in Eq.~(\ref{gb2}), 
justifying the further approximation that $g_{\rm HS}(r)=1$ for $r\ge 5d$.
The accuracy of the variational theory in predicting the equation of state 
has been confirmed by independent comparisons with Monte Carlo simulation 
data~\cite{vRDH99,lu-denton}.

\subsubsection{Grand Potential and Phase Coexistence}\label{Grand}
For a system at fixed temperature, volume, and number of macroions, 
in osmotic equilibrium with a salt reservoir at fixed salt chemical 
potential $\mu_s$, the appropriate thermodynamic potential 
(minimized at equilibrium) is the semi-grand potential,
\begin{equation}
\Omega(T,V,N_m,\mu_s) = F(T,V,N_m,N_s)-\mu_s N_s = -pV+\mu_m N_m,
\label{Omega}
\end{equation}
where $p$ is the bulk pressure and $\mu_m$ is the chemical potential of the
pseudomacroions.  More precisely, $\mu_m$ is the change in free energy 
-- at constant $T$ and $V$ -- upon adding a bare macroion and its $Z/z$ 
neutralizing counterions and $\mu_s$ is the change in free energy 
upon adding a charge-neutral pair of salt ions.  
The semi-grand potential density is then given by
\begin{equation}
\omega(T,n_m,\mu_s) = \Omega/V = f(T,n_m,n_s)-\mu_s n_s = -p+\mu_m n_m,
\label{omega}
\end{equation}
where $f=F/V$ is the total free energy density and $n_s=N_s/V$ is the 
number density of salt ion pairs in the system.  At constant $T$,
the differential relation
\begin{equation}
{\rm d}\Omega(T,V,N_m,\mu_s) = -p{\rm d}V + \mu_m{\rm d}N_m - N_s{\rm d}\mu_s,
\label{dOmega}
\end{equation}
yields the pressure 
\begin{equation}
p = -\left(\frac{\partial \Omega}{\partial V}\right)_{T,N_m,\mu_s} 
= n_m\left(\frac{\partial \omega}{\partial n_m}\right)_{T,\mu_s}-\omega
\label{p1}
\end{equation}
and the macroion chemical potential 
\begin{equation}
\mu_m = \left(\frac{\partial \Omega}{\partial N_m}\right)_{T,V,\mu_s} 
= \left(\frac{\partial \omega}{\partial n_m}\right)_{T,\mu_s}.
\label{mum}
\end{equation}

Equilibrium coexistence of bulk phases requires equality of pressure and of 
chemical potentials (of macroions and salt ions) in the two phases (1 and 2):
\begin{eqnarray}
p^{(1)}&=&p^{(2)} 
\label{coexist-a}
\\
\mu_m^{(1)}&=&\mu_m^{(2)} 
\label{coexist-b}
\\
\mu_s^{(1)}&=&\mu_s^{(2)} = \mu_s^{(r)},
\label{coexist-c}
\end{eqnarray}
where the superscript $(r)$ denotes a reservoir quantity.  Equality of pressure
is equivalent to equality of osmotic pressure, $\Pi=p-p^{(r)}$, {\it i.e.}, 
the difference between the system and reservoir pressures.  The osmotic 
pressure -- a manifestation of the Donnan effect~\cite{hunter} -- vanishes 
in the dilute limit of zero colloid concentration. 

The coexistence conditions have simple geometrical interpretations. 
Equations~(\ref{omega})-(\ref{coexist-c}) describe a common tangent, of slope 
$\mu_m$ and intercept $-p$, to the curve of $\omega(n_m,\mu_s)$ vs. $n_m$ 
(constant $\mu_s$), or equivalently a Maxwell equal-area construction.  
Specifically, the relations
\begin{equation}
\int_1^2{\rm d}\omega=\int_1^2{\rm d}n_m\,\mu_m(n_m,\mu_s)=
\mu_m^{(1)}(n_m^{(2)}-n_m^{(1)})
\label{equal-area1}
\end{equation}
and
\begin{equation}
\int_1^2{\rm d}(\Omega/N_m)=-\int_1^2{\rm d}v_m\,p(v_m,\mu_s)=
-p^{(1)}(v_m^{(2)}-v_m^{(1)}), 
\label{equal-area2}
\end{equation}
with $v_m=V/N_m=1/n_m$, 
imply that constant-$\mu_s$ curves of $\mu_m(n_m,\mu_s)$ vs. $n_m$ 
and of $p(v_m,\mu_s)$ vs. $v_m$ enclose equal areas above and below 
the horizontal lines $\mu_m=\mu_m^{(1)}=\mu_m^{(2)}$ and 
$p=p^{(1)}=p^{(2)}$, respectively.  Changes of curvature sufficient to allow
common-tangent constructions on the semi-grand potential, and equal-area 
constructions on the chemical potential and pressure, imply phase coexistence.  

At low salt concentrations, the salt reservoir behaves as an ideal gas of 
ions, whose pressure and chemical potential are well approximated by 
\begin{equation}
p^{(r)}=2n_s^{(r)}k_BT
\label{pr}
\end{equation}
and
\begin{equation}
\mu_s^{(r)}=2k_BT\ln(n_s^{(r)}\Lambda^3),
\label{musr}
\end{equation}
where $n_s^{(r)}$ is the reservoir number density of pairs of salt ions 
of thermal wavelength $\Lambda$.  Note that $\Lambda$ and $\Lambda_m$ are 
arbitrary, as they contribute to the semi-grand potential only terms 
that are linear in density, which do not affect the coexisting densities. 

The phase diagram is computed as follows.  For a given macroion density $n_m$ 
and salt chemical potential ({\it i.e.}, reservoir salt density $n_s^{(r)}$), 
the system salt density $n_s$ is numerically determined 
[from Eq.~(\ref{coexist-c})] as the solution of
\begin{equation}
\mu_s=\left(\frac{\partial f(n_m,n_s)}{\partial n_s}\right)_{T,n_m}
=2k_BT\ln(n_s^{(r)}\Lambda^3),
\label{ns}
\end{equation}
where $f$ is the total free energy density, the excess part of which is 
given by Eq.~(\ref{gb2}).
In the case of linear response, Eq.~(\ref{ns}) can be expressed in a somewhat 
more practical form by separating out and analytically evaluating the dominant 
volume energy contribution.  Substituting Eqs.~(\ref{E03}) and (\ref{Ftot})
into Eq.~(\ref{ns}) then yields
\begin{equation}
\beta\mu_s= \ln[(n_c+n_s)\Lambda^3]+\ln(n_s\Lambda^3)
-\frac{Z\kappa\lambda_B}{2(1+\kappa a)^2}\frac{n_c}{n_{\mu}}
+\left(\frac{n_c}{n_{\mu}}\right)^2
+\beta\left(\frac{\partial f_{\rm ex}(n_m,n_s)}{\partial n_s}\right)_{n_m}.
\label{mus}
\end{equation}
The pressure and macroion chemical potential are next computed from 
Eqs.~(\ref{p1}) and (\ref{mum}).  Finally, the macroion and salt densities
are varied to satisfy the remaining coexistence conditions 
[Eqs.~(\ref{coexist-a}) and (\ref{coexist-b})].

\section{Results and Discussion}\label{Results}
To investigate the influence of nonlinear microion screening on the phase 
behavior of deionized charged colloids, the variational theory 
(Sec.~\ref{PhaseBehavior}) is used to compute the semi-grand potential, 
taking as input the effective interactions predicted by response theory 
(Sec.~\ref{Interactions}).  By performing a coexistence analysis and comparing 
the phase diagrams that result from linear and first-order nonlinear 
interactions, leading-order nonlinear effects are quantified.  For simplicity, 
effective three-body interactions are here neglected, since these are always 
attractive~\cite{denton04} and thus would only promote phase separation.  
In this way, we isolate the main nonlinear corrections to the volume energy 
and effective pair potential and assess their impact on phase behavior.

\begin{figure}
\includegraphics[width=0.75\columnwidth]{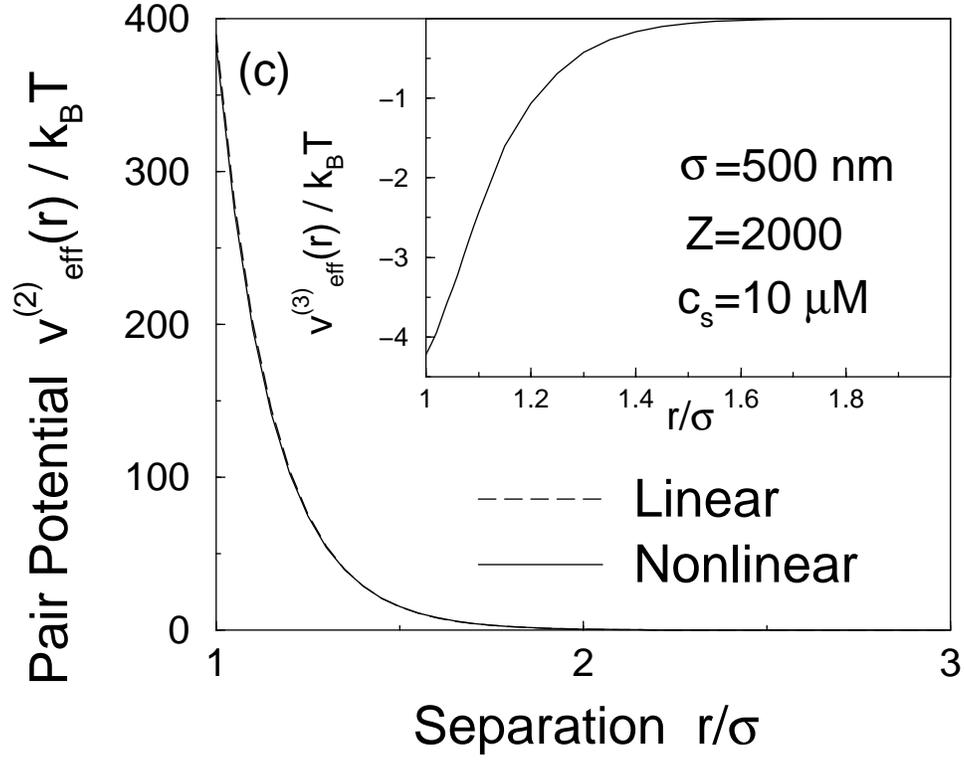} 
\caption{\label{vr} 
Effective pair potential $v^{(2)}_{\rm eff}(r)$ vs. center-to-center separation
$r$ for fixed colloid volume fraction $\eta=0.05$ and various combinations of
macroion diameter $\sigma$, effective valence $Z$, and system salt 
concentration $c_s$:
(a) $\sigma=100$ nm, $Z=500$, $c_s=50~\mu$M;
(b) $\sigma=266$ nm, $Z=1217$, $c_s=10~\mu$M;
(c) $\sigma=500$ nm, $Z=2000$, $c_s=10~\mu$M.
Solid (dashed) curves are predictions of nonlinear (linear) response theory.
Insets show corresponding effective triplet potentials $v^{(3)}_{\rm eff}(r)$ 
for three macroions arranged in an equilateral triangle of side length $r$.
}
\end{figure}

Numerical results are presented for the case of room-temperature aqueous 
suspensions ($\lambda_B=0.72$ nm) and monovalent counterions ($z=1$).
For several choices of macroion radius $a$, the effective macroion valence $Z$ 
is set near the threshold for charge renormalization~\cite{alexander84}, 
$Z\sim O(10)(a/\lambda_B)$.  Figure~\ref{vr} illustrates the effective pair 
and triplet potentials vs.  macroion separation, with linear and nonlinear 
screening, for various sets of system parameters.  The particular case of
($\sigma=266$ nm, $Z=1217$) is included to permit direct comparison with
ref.~\cite{vRDH99}.  While nonlinear screening 
generally softens repulsive pair interactions, the correction is relatively 
minor for the selected macroion diameters and valences.  The effective triplet 
potential, shown for an equilateral triangle arrangement of three macroions, 
is always attractive and decays rapidly with increasing separation.
In passing, we note that the triplet interactions that arise within response 
theory~\cite{denton04} differ in definition from their counterparts in 
Poisson-Boltzmann theory~\cite{russ02,hynninen04}.

Figures~\ref{pressure-lin} and \ref{pressure} present predictions for the 
osmotic pressure $\Pi$ (equation of state) vs. volume fraction $\eta$ at 
fixed reservoir salt concentration $c_s^{(r)}$ (or salt chemical potential
$\mu_s$).  The variation of $\Pi$ with $\eta$ is a diagnostic of 
thermodynamic stability, a negative slope signaling instability toward 
phase separation (see below).  Figure~\ref{pressure-lin} illustrates that, 
within the linearized theory, the system becomes unstable below a certain 
critical salt concentration.  Figure~\ref{pressure} demonstrates the 
sensitivity of the osmotic pressure to nonlinear screening, which originates 
mainly from the nonlinear correction to the volume energy.  

\begin{figure}
\includegraphics[width=0.75\columnwidth]{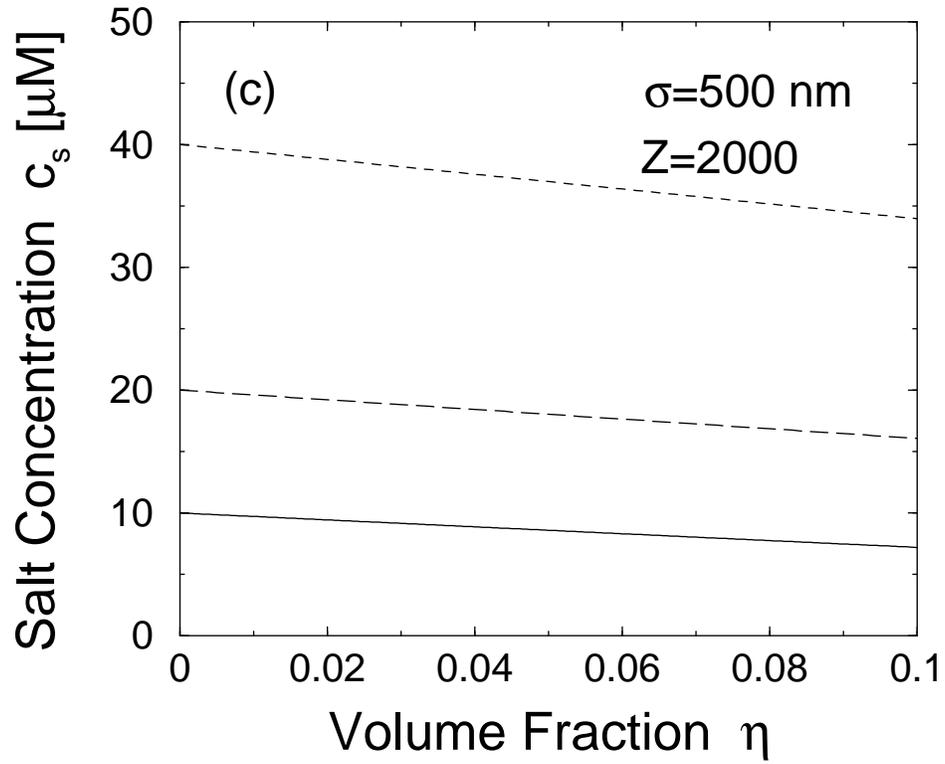}
\caption{\label{salt-lin} 
Linear-screening predictions for system salt concentration $c_s$ 
[$\mu$mol/liter] vs. colloid volume fraction $\eta$ for same system 
parameters as in Fig.~\ref{vr} at various fixed salt chemical potentials.
Respective reservoir salt concentrations are given by intersections of 
curves with $\eta=0$ axis.  
}
\end{figure}

\begin{figure}
\includegraphics[width=0.75\columnwidth]{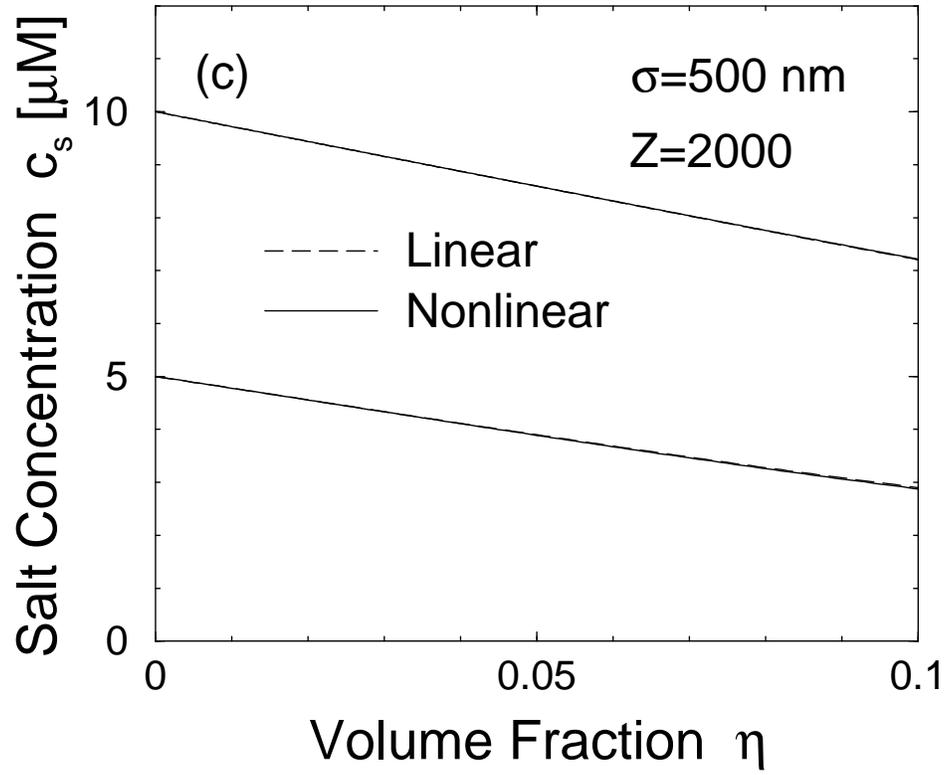}
\caption{\label{salt} 
Linear- and nonlinear-screening predictions for system salt concentration 
$c_s$ [$\mu$mol/liter] vs. colloid volume fraction $\eta$ for same system 
parameters as in Fig.~\ref{vr} and at two fixed salt chemical potentials.
Respective reservoir salt concentrations are given by intersections of 
curves with $\eta=0$ axis.  
Solid (dashed) curves are predictions of linear (nonlinear) response theory.
}
\end{figure}

Figures~\ref{salt-lin} and \ref{salt} present the corresponding system salt 
concentration $c_s$ (in $\mu$mol/liter) vs. volume fraction (at fixed $\mu_s$).
The monotonic decrease of $c_s$ with increasing $\eta$ follows from 
Eq.~(\ref{mus}) and stems from an interplay between salt entropy and 
salt-macroion interactions.  
Entropy and excluded-volume interactions alone would give a simple linear 
decline, $c_s=(1-\eta)c_s^{(r)}$, with a slope of $-c_s^{(r)}$.  However, 
salt-macroion electrostatic interactions tend to expel salt from the system, 
steepening the decline, while maintaining an approximate linear dependence 
over a considerable range of $\eta$.  As illustrated in Fig.~\ref{salt}, 
nonlinear screening, which modifies the state dependence of the effective 
interactions, tends to lower the system salt concentration.

\begin{figure}
\includegraphics[width=0.75\columnwidth]{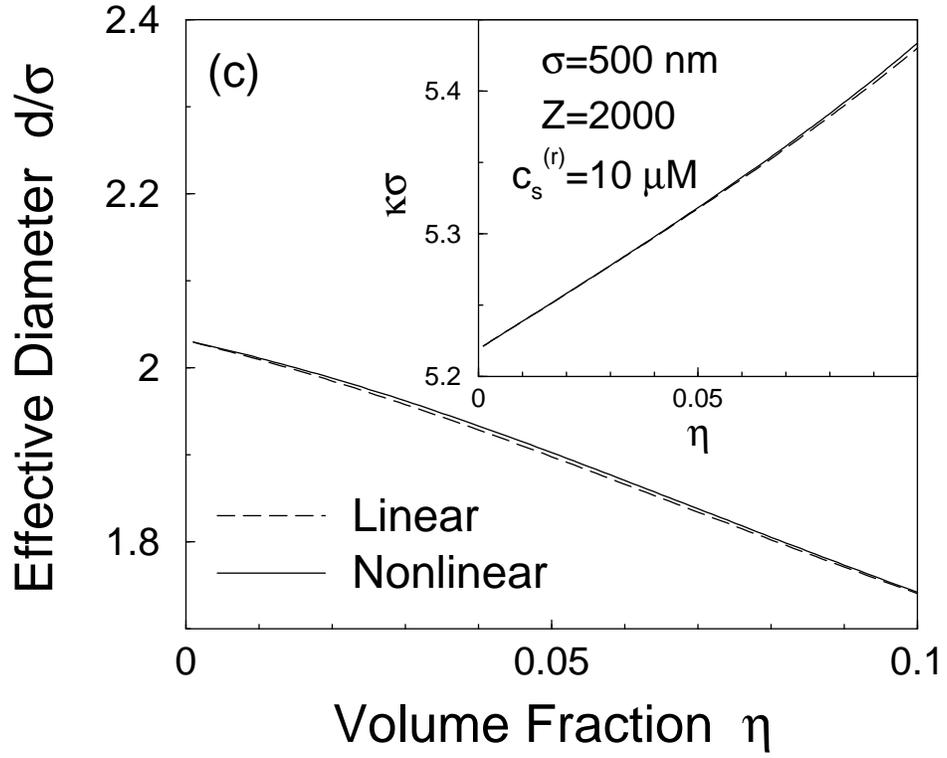}
\caption{\label{kappad} 
Effective hard-sphere diameter $d$ (units of macroion diameter $\sigma$)
and Debye screening constant $\kappa$ (inset) vs. colloid volume fraction 
$\eta$, at fixed reservoir salt concentration $c_s^{(r)}$, for 
(a) $\sigma=100$ nm, $Z=500$, $c_s^{(r)}=50~\mu$M;
(b) $\sigma=266$ nm, $Z=1217$, $c_s^{(r)}=10~\mu$M;
(c) $\sigma=500$ nm, $Z=2000$, $c_s^{(r)}=10~\mu$M.
Solid (dashed) curves are predictions of nonlinear (linear) response theory.
}
\end{figure}

Figure~\ref{kappad} typifies the monotonic decrease of the effective 
hard-sphere diameter $d$, and increase of the Debye screening constant $\kappa$,
with increasing volume fraction at fixed $\mu_s$.  Nonlinear screening 
evidently reduces both $d$ and $\kappa$.  For the chosen parameters, 
the reduction appears modest, but is significant, given the sensitivity 
of the free energy to these parameters.

Figures~\ref{pressure-lin}, \ref{pressure}, and \ref{cplin} illustrate that,
for sufficiently high macroion valence and low salt concentration, van der 
Waals loops emerge in the equation of state at fixed $\mu_s$ -- a direct 
signature of phase instability.  
The maximimum and minimum in the curve of osmotic pressure vs. volume fraction
mark the vapor and liquid spinodal densities, respectively, between which the 
compressibility is negative and the uniform fluid is unstable with respect to 
phase separation [Fig.~\ref{cplin}(a)].  Correspondingly, an equal-area 
construction on the curve of osmotic pressure vs. inverse volume fraction 
[Fig.~\ref{cplin}(b)], or of chemical potential vs.  volume fraction 
[Fig.~\ref{cplin}(c)], yields the densities of the coexisting vapor and liquid 
phases.  A scan over reservoir salt concentration (salt chemical potential)  
traces out the spinodal and binodal (coexistence) curves in the phase diagram.

Figure~\ref{phase-diagram} presents the resulting fluid phase diagrams for 
highly deionized suspensions as predicted by variational theory with both 
linear and nonlinear effective interactions as input.  In each case, above a 
critical salt concentration, the uniform fluid is thermodynamically stable.  
Below the critical point, the fluid separates into macroion-rich (liquid) 
and macroion-poor (vapor) bulk phases, the salt concentration playing a role 
analogous to temperature in the liquid-vapor separation of a simple 
one-component fluid.  For the parameter regime investigated here, the density 
of the liquid phase is found to be always well below the threshold for 
freezing, estimated from the hard-sphere freezing criterion, 
$\eta(d/\sigma)^3\sim0.49$, with the charged colloids approximated 
as neutral hard spheres of effective diameter $d$.

The tie lines in the phase diagrams of Fig.~\ref{phase-diagram} join 
corresponding points on the liquid and vapor binodals (and spinodals) and, 
if extended, intersect the $\eta=0$ axis at the respective reservoir salt 
concentrations.  The fact that the tie lines all have essentially the same 
slope, independent of reservoir salt concentration, is a physical consequence 
of strong salt-macroion electrostatic interactions, as described by 
Eq.~(\ref{mus}).  The influence of nonlinear response on the tie-line slopes 
is negligible for the parameters here investigated.

The predicted phase separation of charged colloids is remarkable, considering
that simple one-component systems, interacting via purely repulsive pair 
potentials, exhibit only a single fluid phase.  
Within the present theoretical framework, 
phase instability at low salt concentrations is driven by the strong density 
dependence of the effective interactions, chiefly the one-body volume energy 
in deionized suspensions.  It should be emphasized that because the colloid 
and salt concentrations vary between the two phases, the density-dependent 
effective interactions also differ in the two phases.  

The unusual phase separation can be understood, more fundamentally, as the 
result of a classic competition between entropy and energy.  On one side 
of the balance, favoring a stable uniform fluid, are the configurational 
entropies of all ions, represented by the ideal-gas terms in Eqs.~(\ref{E03})
and (\ref{Ftot}), and the positive potential energy of macroion pair repulsion.
On the other side is the (density-dependent) negative potential energy of 
macroion-counterion attraction [second term on the right side of 
Eq.~(\ref{E03})], which favors a concentrated phase with counterions 
localized around, and thus strongly attracted to, the macroions.  

Within the ``entropy vs. energy" view,
the sensitivity of phase behavior to salt concentration becomes clearer.
At salt concentrations low enough that screening is counterion-dominated and 
screening lengths are relatively long, the counterion distribution is so 
diffuse that counterion-macroion attraction is too weak to drive macroion 
aggregation.  With increasing salt concentration, the screening length 
shortens, the counterions become more localized around the macroions, and
counterion-macroion attraction may -- for sufficiently high macroion valence
-- overcome configurational entropy and macroion pair repulsion to drive 
phase separation.  The resulting concentrated phase is energetically favored, 
the counterions being closer on average to the macroions, but entropically 
disfavored, since the microions (excluded by macroion cores) must occupy 
a smaller free volume.  On the other hand, the dilute phase is energetically 
disfavored, the counterions tending to roam farther from the macroions, but is 
entropically favored, since the microions can explore a larger free volume.  
At salt concentrations high enough that screening is salt-dominated, the 
salt-ion entropy overwhelms the counterion-macroion interaction energy
in the free energy and prevents macroion aggregation.

Thermodynamic phase behavior qualitatively similar to that depicted in 
Fig.~\ref{phase-diagram} has been predicted before~\cite{vRDH99,warren00}.  
Compared with the results of van Roij {\it et al.}~\cite{vRDH99}, based on 
essentially the same variational theory for free energies, but a linearized 
density-functional theory for effective interactions, the present theory
predicts a somewhat larger unstable area in the phase diagram.  This 
quantitative discrepancy results mainly from different treatments of 
excluded-volume effects in the two approaches.  In particular, the 
excluded-volume correction to the screening constant in response theory 
[$1/(1-\eta)$ factor in Eq.~(\ref{kappa})] enhances microion screening and 
promotes phase instability.

\section{Summary and Conclusions}\label{Conclusions}
\begin{figure}
\includegraphics[width=0.75\columnwidth]{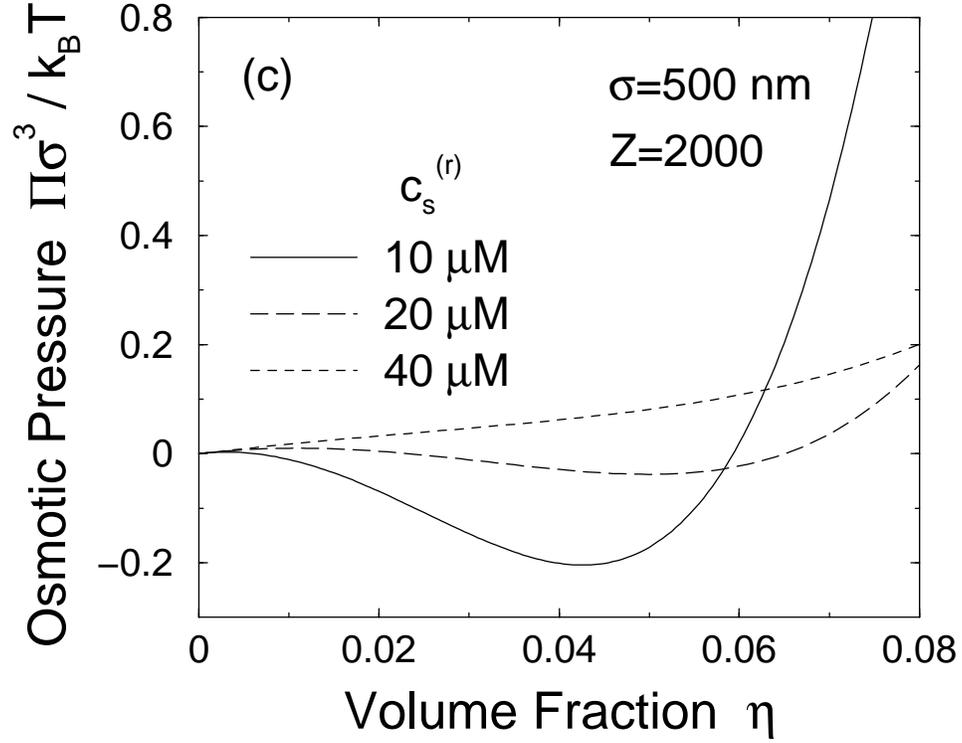} 
\caption{\label{pressure-lin} 
Linear-screening predictions for osmotic pressure $\Pi$ (in reduced units) 
vs. colloid volume fraction $\eta$ for same combinations of macroion diameter 
$\sigma$ and valence $Z$ as in Fig.~\ref{vr} and various fixed reservoir 
salt concentrations $c_s^{(r)}$:
(a) $\sigma=100$ nm, $Z=500$, $c_s^{(r)}=100, 200, 400~\mu$M;
(b) $\sigma=266$ nm, $Z=1217$, $c_s^{(r)}=40, 80, 160~\mu$M;
(c) $\sigma=500$ nm, $Z=2000$, $c_s^{(r)}=10, 20, 40~\mu$M.
}
\end{figure}

\begin{figure}
\includegraphics[width=0.75\columnwidth]{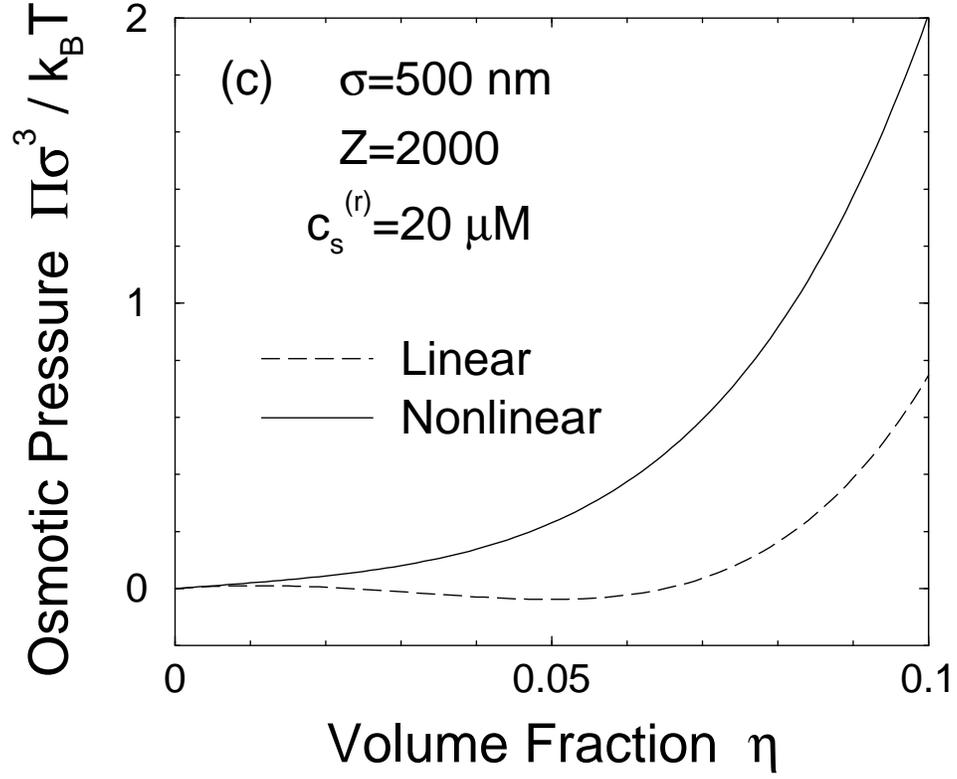} 
\caption{\label{pressure} 
Osmotic pressure $\Pi$ (in reduced units) vs. colloid volume fraction $\eta$ 
for same combinations of macroion diameter $\sigma$ and valence $Z$ as in
Fig.~\ref{vr} and fixed reservoir salt concentration $c_s^{(r)}$:
(a) $\sigma=100$ nm, $Z=500$, $c_s^{(r)}=200~\mu$M;
(b) $\sigma=266$ nm, $Z=1217$, $c_s^{(r)}=80~\mu$M;
(c) $\sigma=500$ nm, $Z=2000$, $c_s^{(r)}=20~\mu$M.
Solid (dashed) curves are predictions of nonlinear (linear) response theory.
}
\end{figure}

\begin{figure}
\includegraphics[width=0.75\columnwidth]{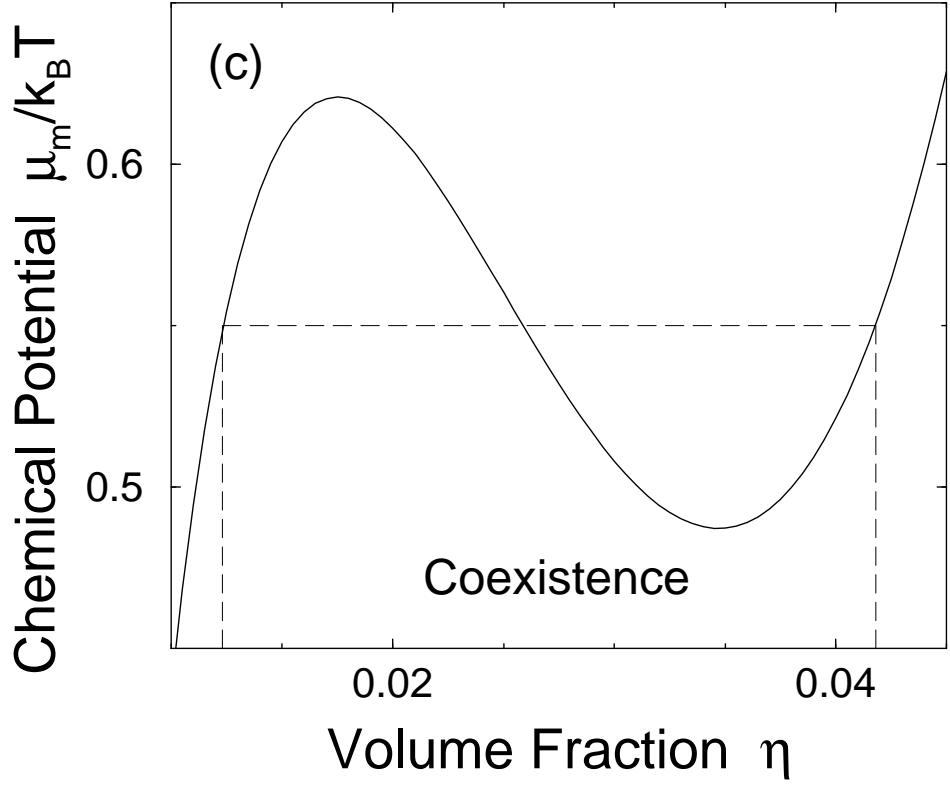}
\caption{\label{cplin} 
Linear-screening prediction for (a) osmotic pressure $\Pi$ vs. colloid volume 
fraction $\eta$, (b) $\Pi$ vs. $1/\eta$, and (c) colloid chemical potential 
$\mu_m$ (shifted by arbitrary constant) vs. $\eta$ for macroion diameter 
$\sigma=100$ nm, valence $Z=500$, and reservoir salt concentration 
$c_s^{(r)}=350~\mu$M.  
In panels (a) and (b), dotted vertical lines at maximum and minimum of $\Pi$ 
indicate spinodal densities at boundaries of unstable region.  In panels 
(b) and (c), dashed vertical lines indicate coexisting densities on the
fluid binodal, illustrating the Maxwell equal-area construction.  
}
\end{figure}

\begin{figure}
\includegraphics[width=0.75\columnwidth]{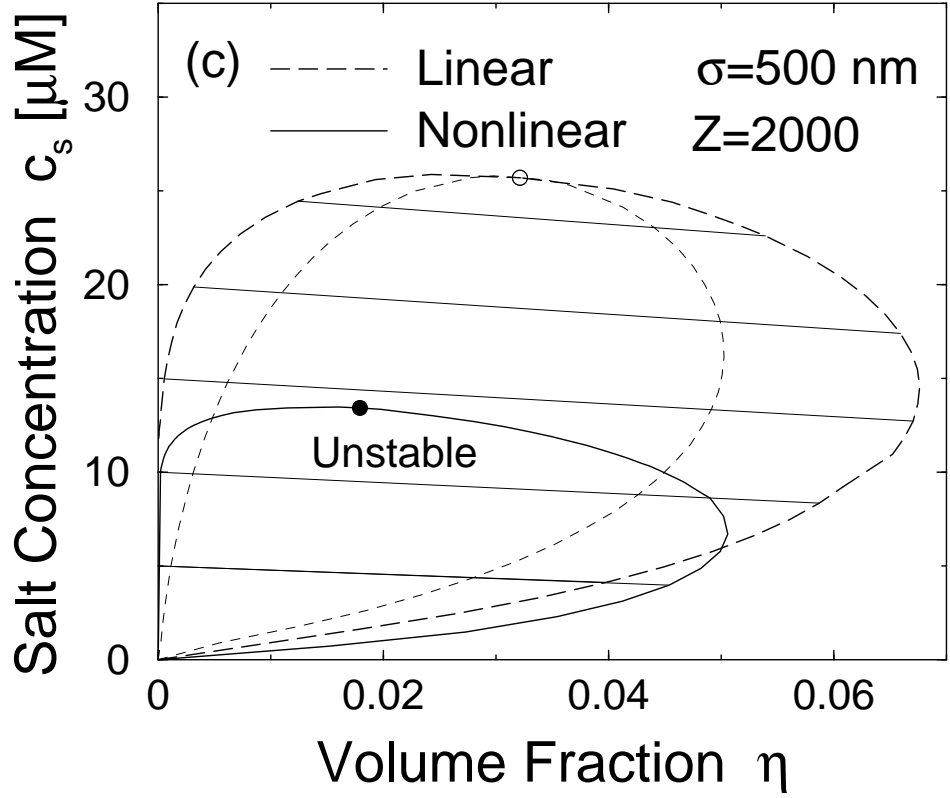} 
\caption{\label{phase-diagram} 
Fluid phase diagrams for aqueous suspensions of charged colloids at room 
temperature ($\lambda_B=0.72$ nm) with monovalent microions and various 
macroion diameters and effective valences:
(a) $\sigma=100$ nm, $Z=500$; 
(b) $\sigma=266$ nm, $Z=1217$; 
(c) $\sigma=500$ nm, $Z=2000$.
Solid (long-dashed) curves represent predictions for binodals from 
nonlinear (linear) response theory.  Short-dashed curves represent 
predictions for spinodals (linear response only).  
Circular symbols denote critical points.
Tie lines join corresponding points on liquid and vapor branches of binodals.
}
\end{figure}

In summary, we have investigated the controversial issue of phase separation 
in deionized charge-stabilized colloidal suspensions by inputting effective 
electrostatic interactions from response theory into free energies from a 
thermodynamic variational theory.  By considering both linear and first-order
nonlinear approximations for the effective pair potential and one-body volume 
energy, we have systematically assessed the influence of nonlinear screening
on phase behavior.  A coexistence analysis results in osmotic pressures 
[Figs.~\ref{pressure-lin}, \ref{pressure}, and \ref{cplin}] and phase diagrams 
[Fig.~\ref{phase-diagram}] that clearly exhibit thermodynamic instability 
towards phase separation for sufficiently high macroion effective valences 
and low salt concentrations.  

For macroion sizes and effective valences within limits established by charge 
renormalization considerations, first-order nonlinear corrections to the 
effective interactions are relatively weak and can either enhance or diminish 
stability of the uniform fluid phase, depending on system parameters.  
In general, the higher the macroion surface charge density, the higher the 
critical salt concentration and the larger the area of the unstable region 
in the phase diagram.  Our main conclusion is that, within the present model,
nonlinear screening appears not to suppress phase separation of deionized 
suspensions, contradicting conclusions drawn from previous 
studies~\cite{vongrunberg01,deserno02,tamashiro03,levin03} and raising hope 
that a similar phenomenon may yet be observed in simulations of the 
primitive model.

In closing, three key approximations of the present approach deserve to be
highlighted for further scrutiny.  First, the neglect of higher-order nonlinear
corrections to the effective interactions presumes that nonlinear effects are 
strongest at the one- and two-body levels.  The finding that first-order 
nonlinear corrections do not qualitatively alter fluid phase behavior suggests 
that higher-order corrections are unlikely to have drastic consequences -- 
for example, suppression of phase separation.  Furthermore, the presumption of
weak many-body effective interactions is consistent with the dominance of 
the volume energy in effective one-component models of 
simple metals~\cite{Pethick70,Brovman70,Singh73,Rasolt75,Louis98}, but should 
be further checked for charged colloids.  Second, the mean-field approximation
for the response functions of the microion plasma assumes weakly correlated 
microions.  Although usually considered reasonable for monovalent microions, 
this assumption can and should be checked by more accurately modeling the 
structure of the microion plasma.  Finally, the assumption of fixed macroion 
valence neglects the dependence of the effective valence on colloid and salt 
densities.  This interesting issue of coupling between the effective macroion 
charge and phase behavior is being examined by means of charge renormalization 
theory and will be the subject of a future paper.


\begin{acknowledgments} 
This work was supported by the National Science Foundation under Grant 
Nos.~DMR-0204020 and EPS-0132289.
\end{acknowledgments} 





\newpage




\unitlength1mm

\begin{figure}
\includegraphics[width=0.75\columnwidth]{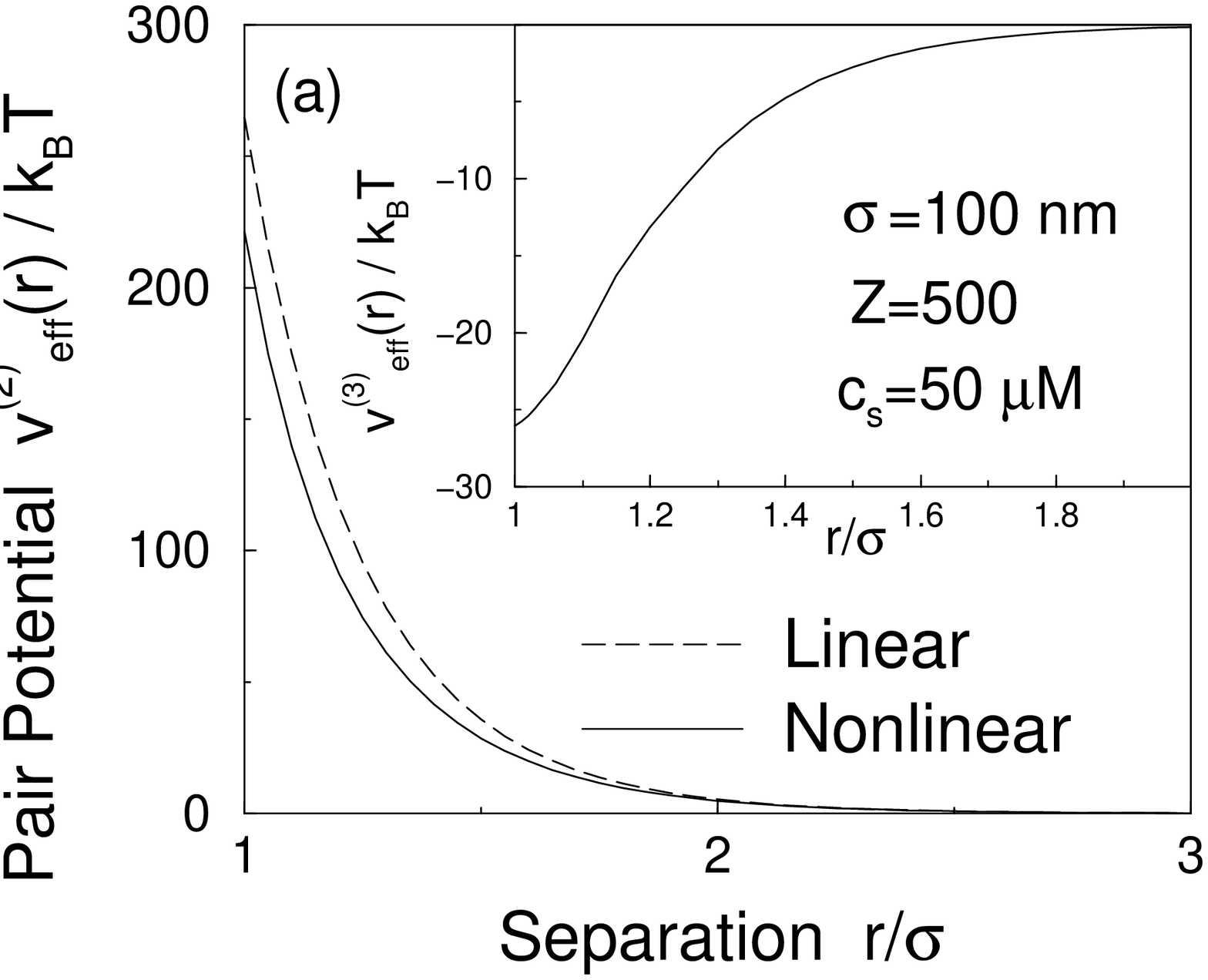} \\[2ex]
\includegraphics[width=0.75\columnwidth]{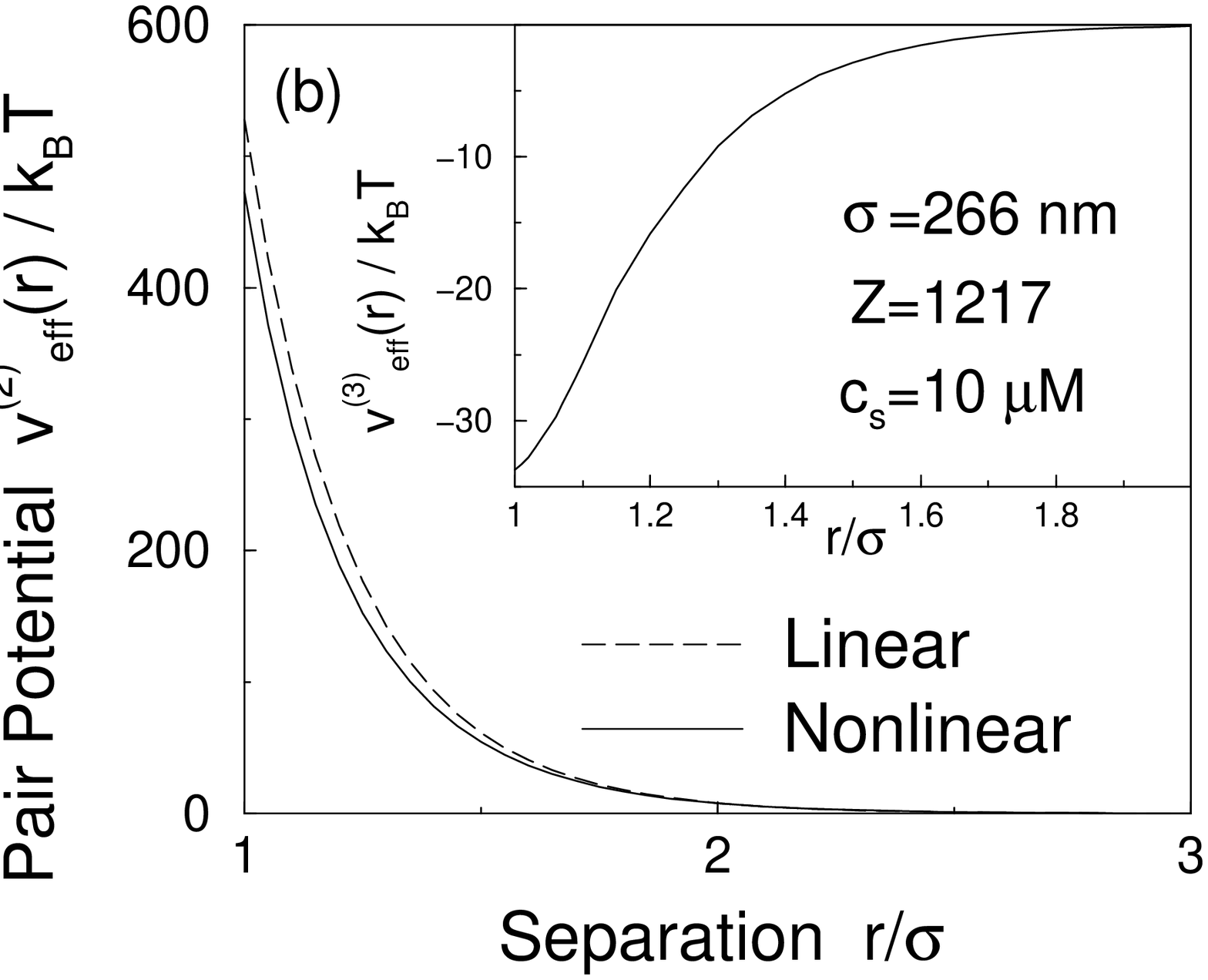} 
\end{figure}

\begin{figure}
\includegraphics[width=0.75\columnwidth]{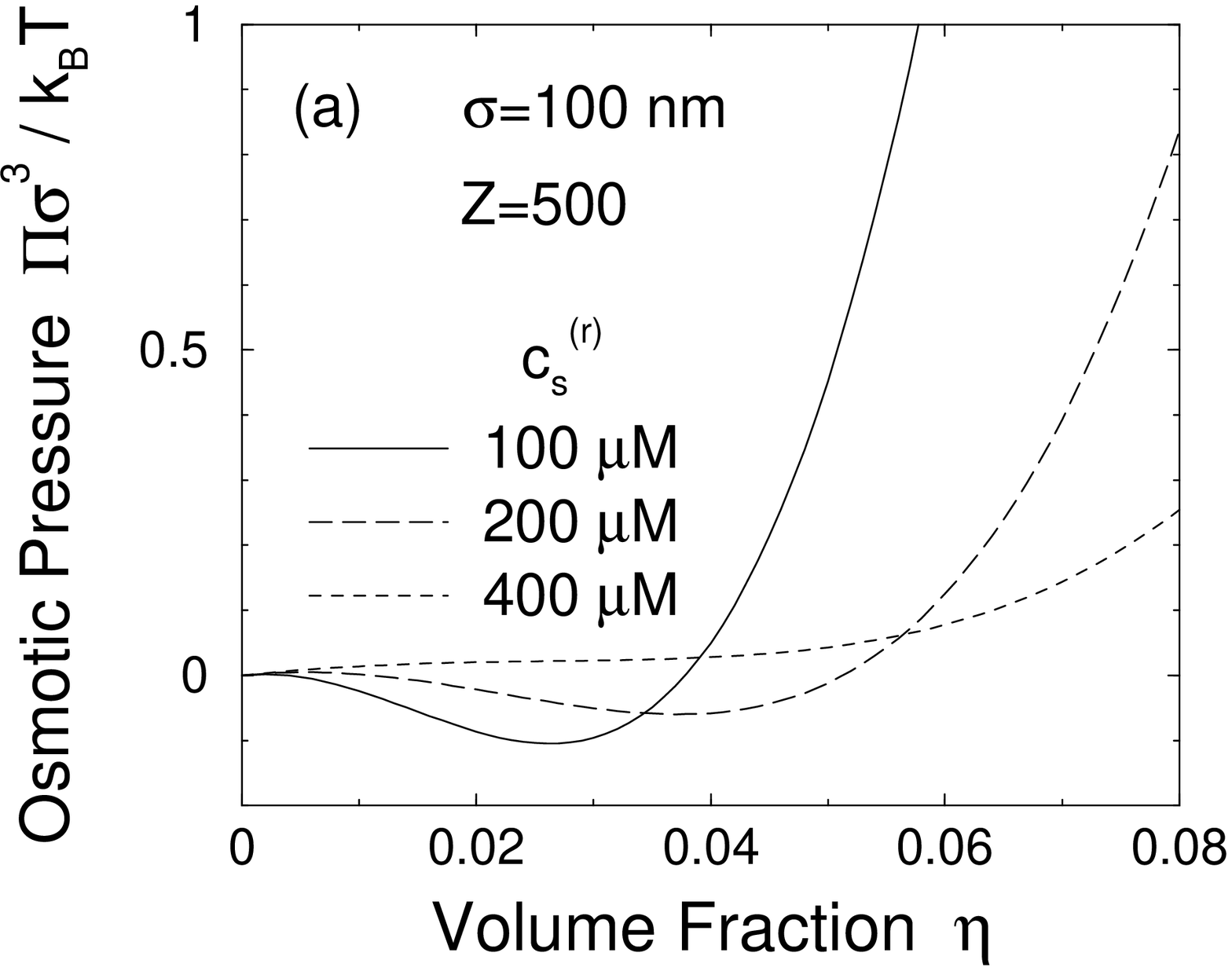} \\[2ex]
\includegraphics[width=0.75\columnwidth]{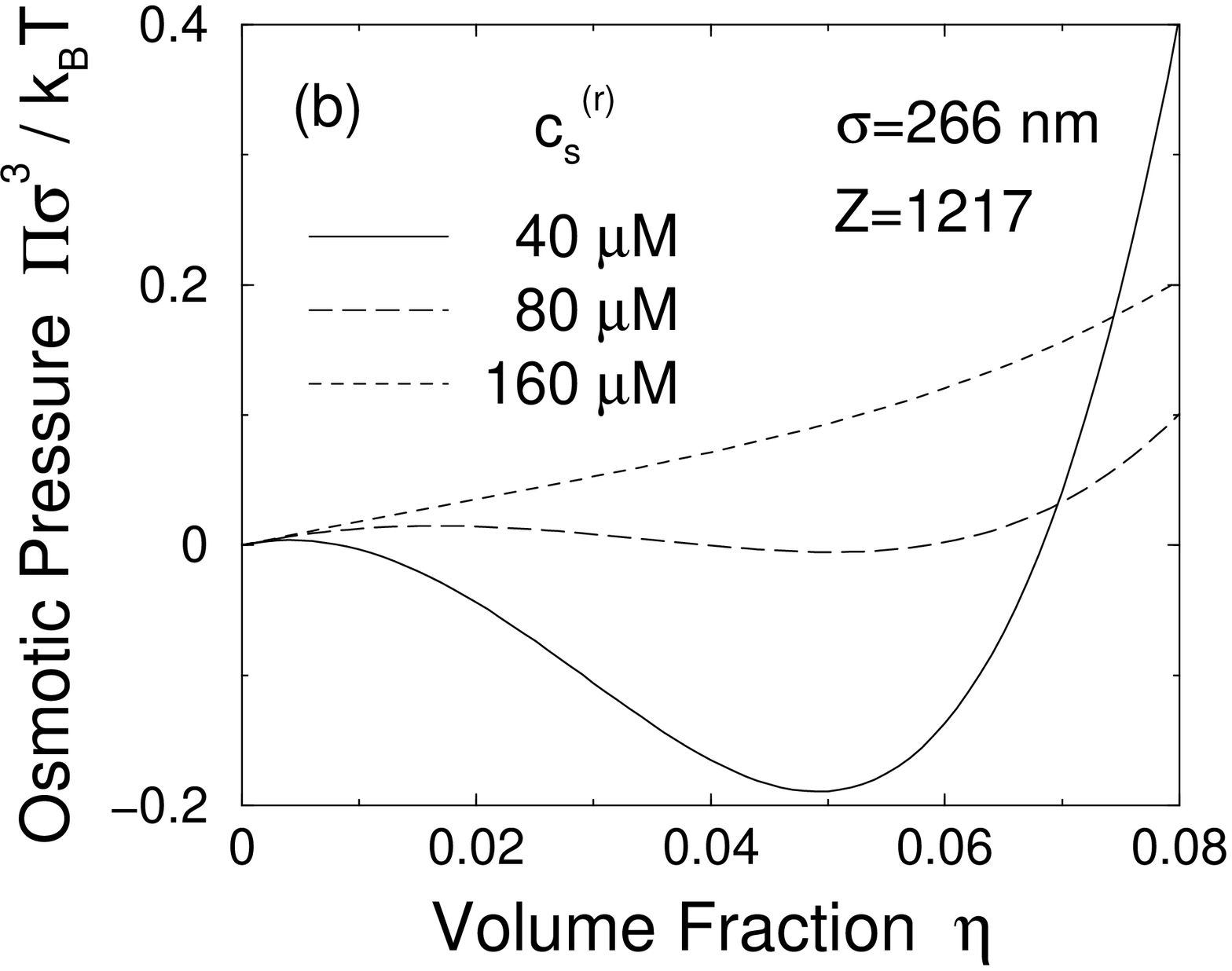} 
\end{figure}

\begin{figure}
\includegraphics[width=0.75\columnwidth]{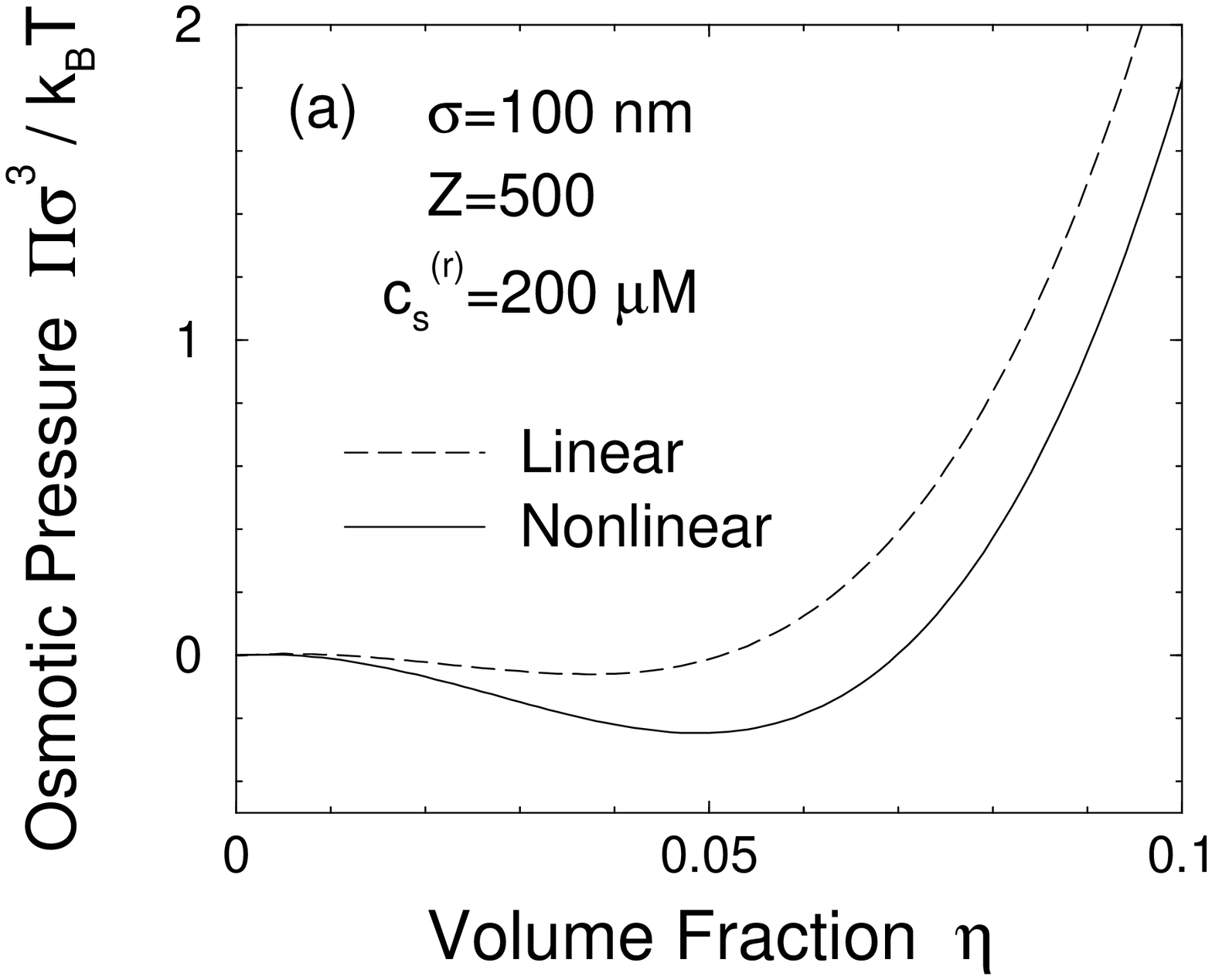} \\[2ex]
\includegraphics[width=0.75\columnwidth]{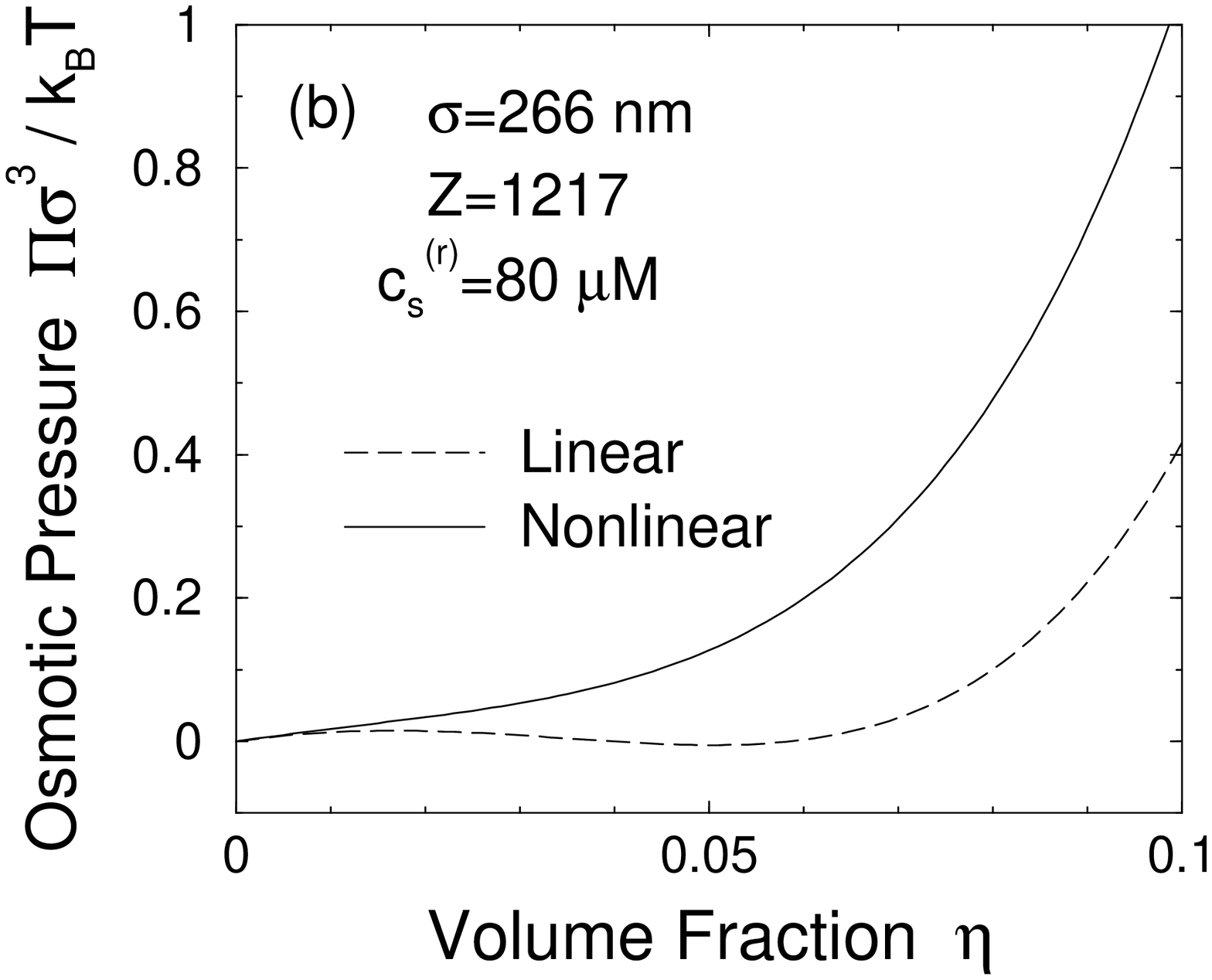} 
\end{figure}

\begin{figure}
\includegraphics[width=0.75\columnwidth]{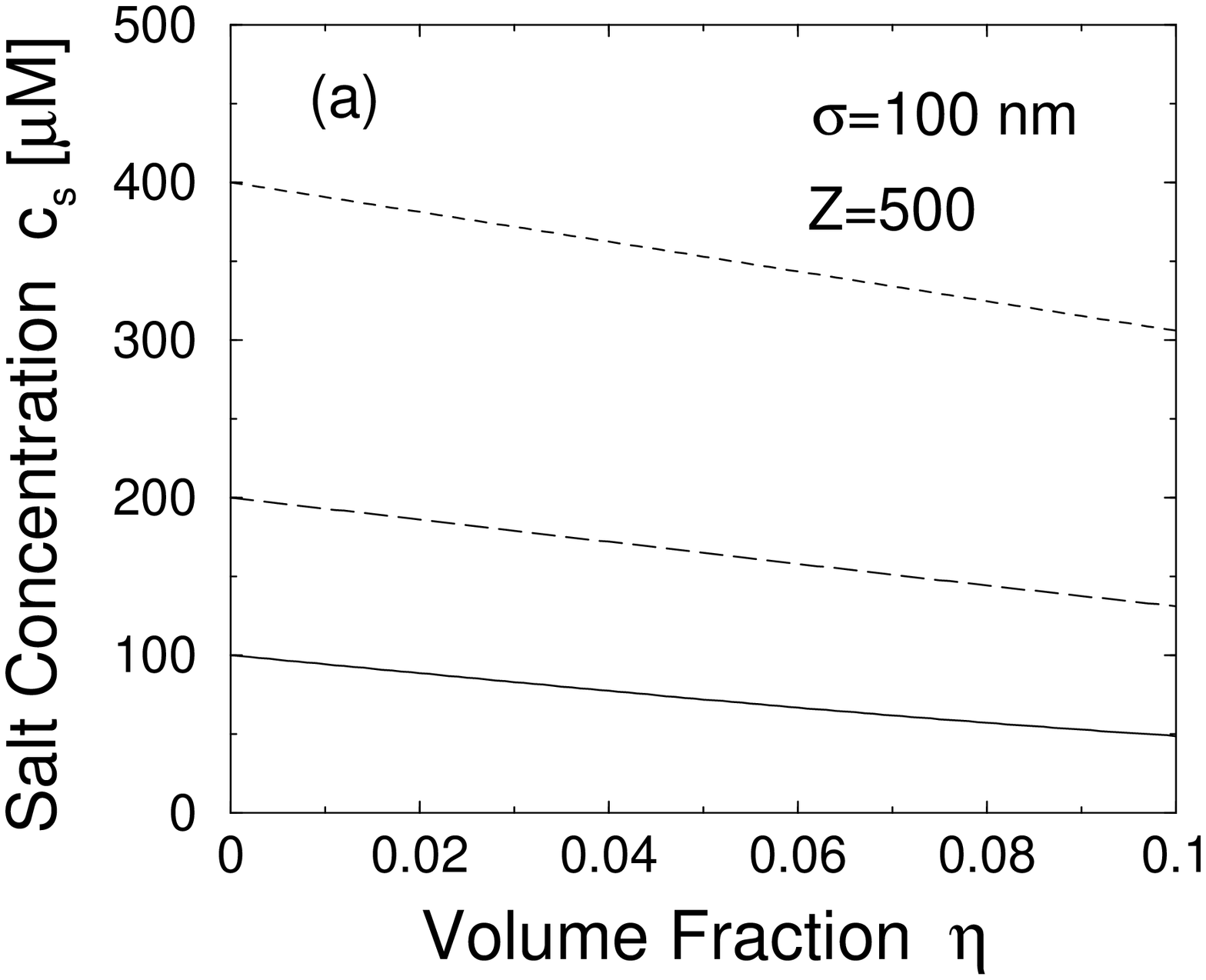} \\[2ex]
\includegraphics[width=0.75\columnwidth]{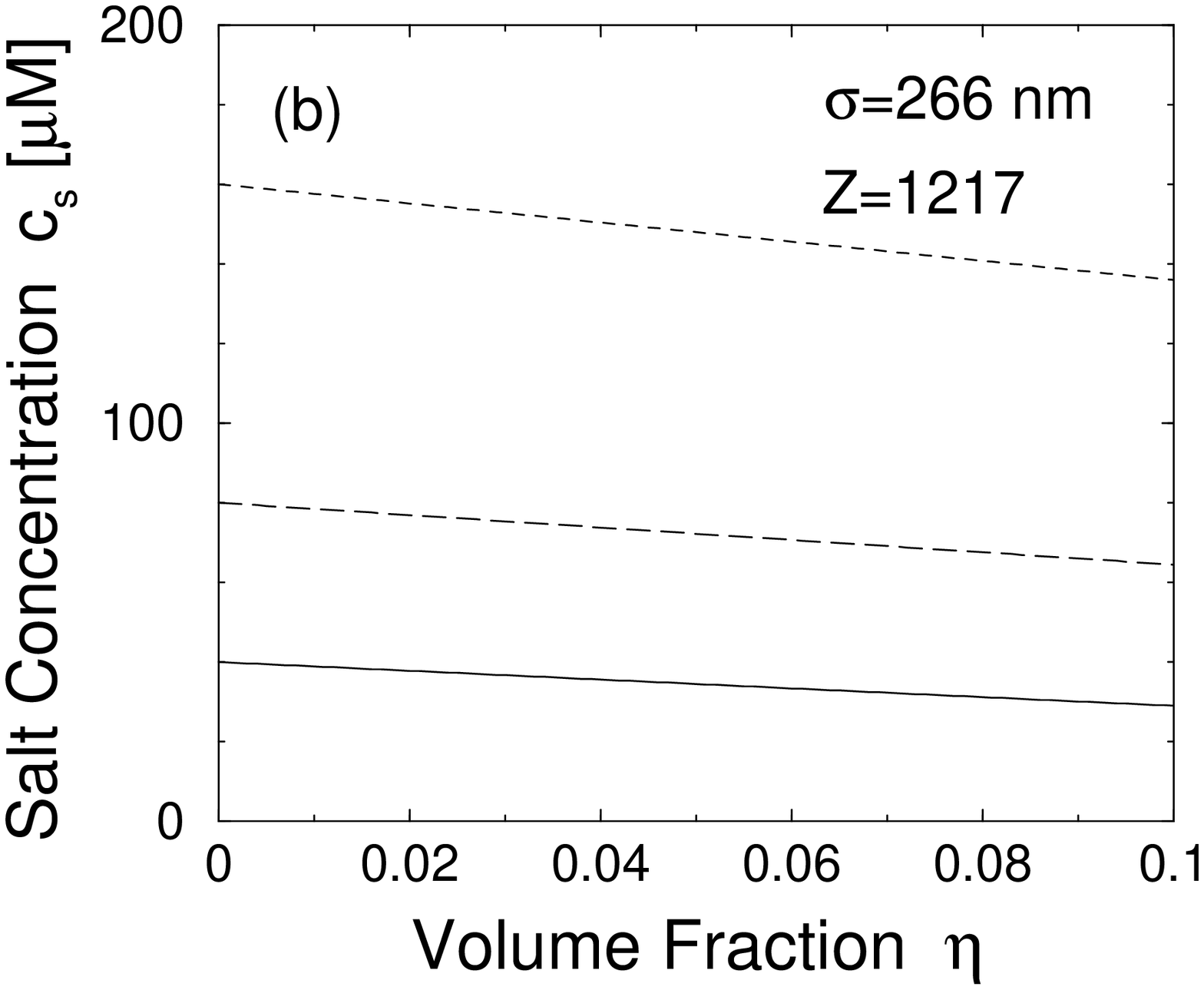}
\end{figure}

\begin{figure}
\includegraphics[width=0.75\columnwidth]{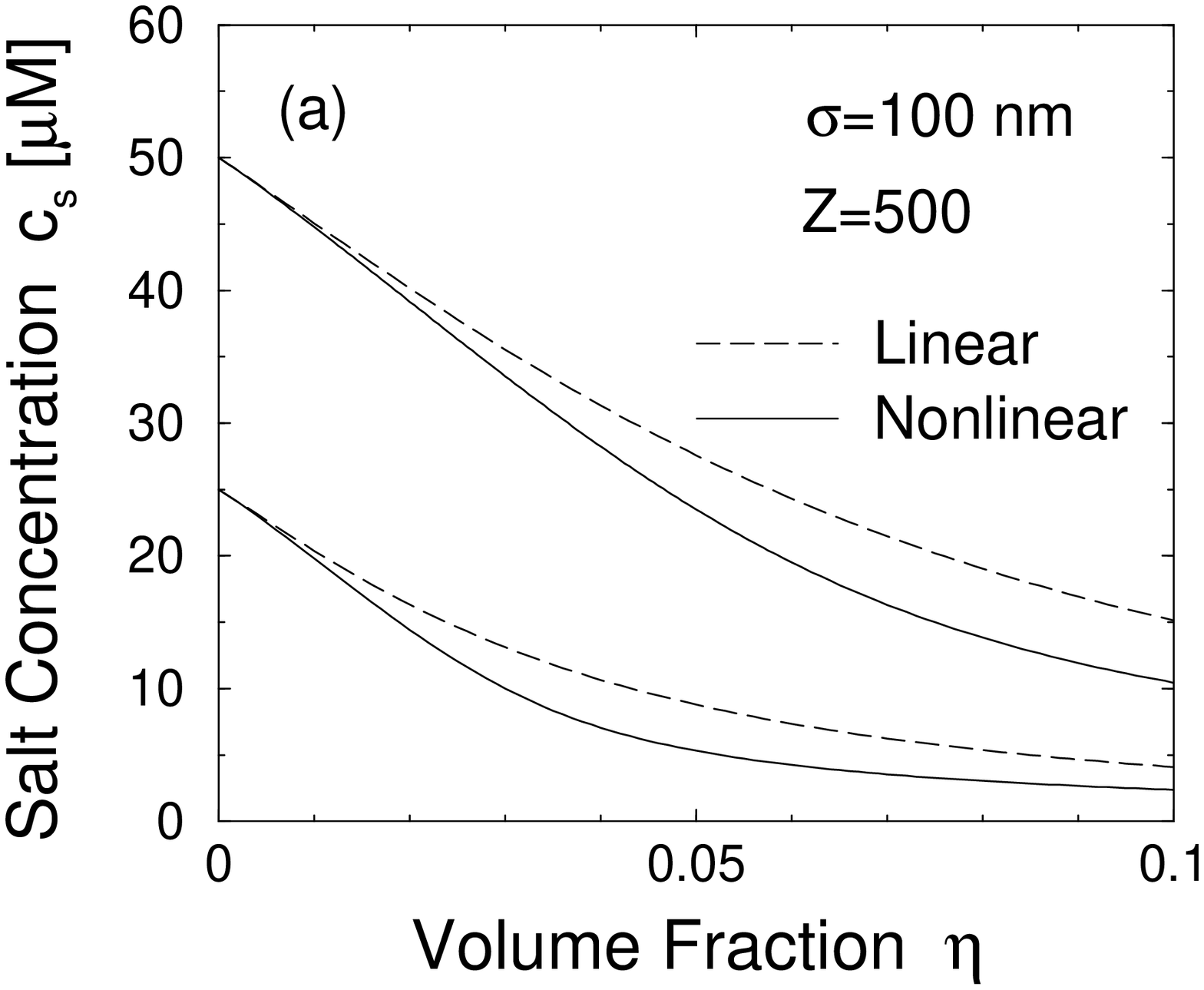} \\[2ex]
\includegraphics[width=0.75\columnwidth]{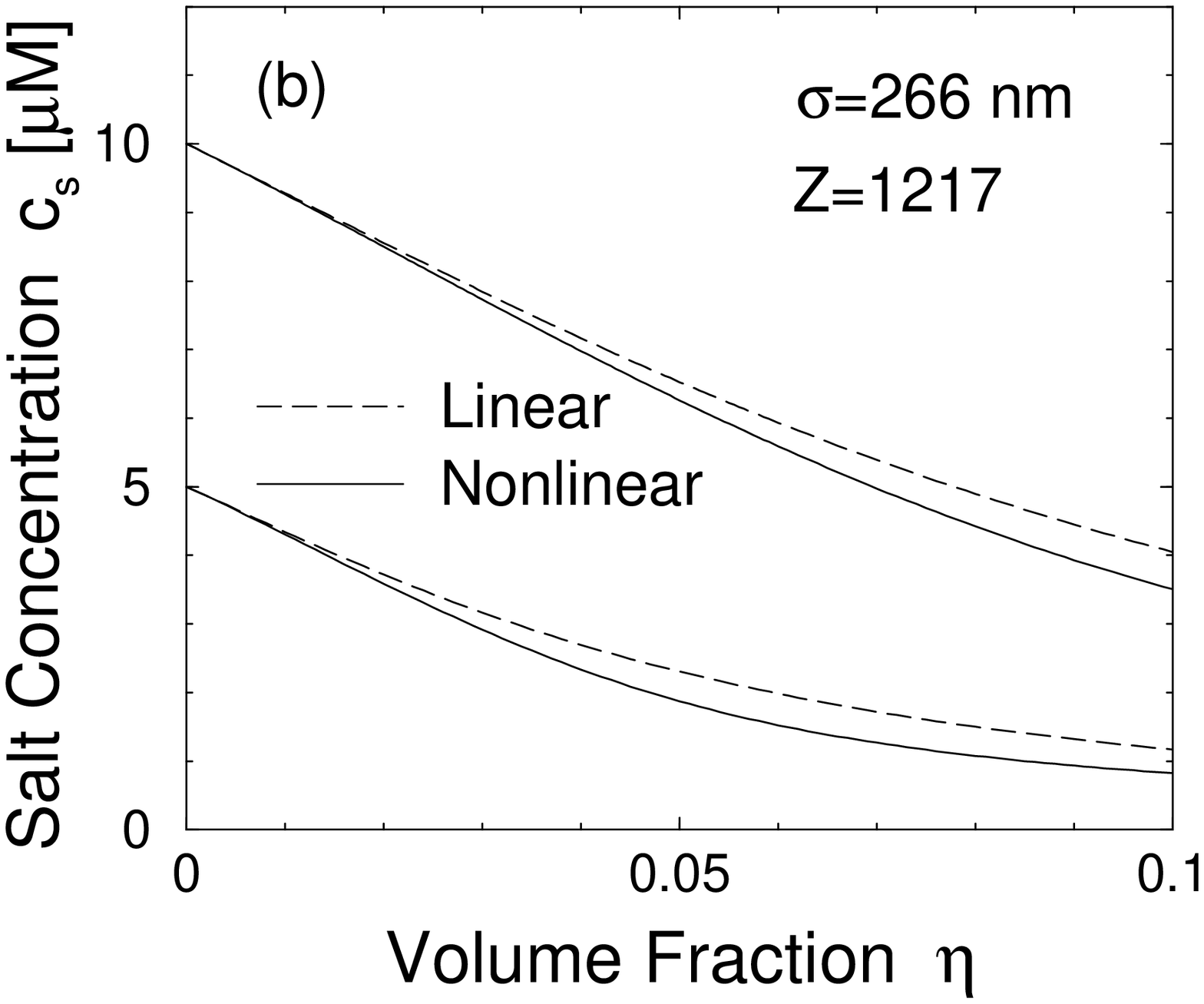}
\end{figure}

\begin{figure}
\includegraphics[width=0.75\columnwidth]{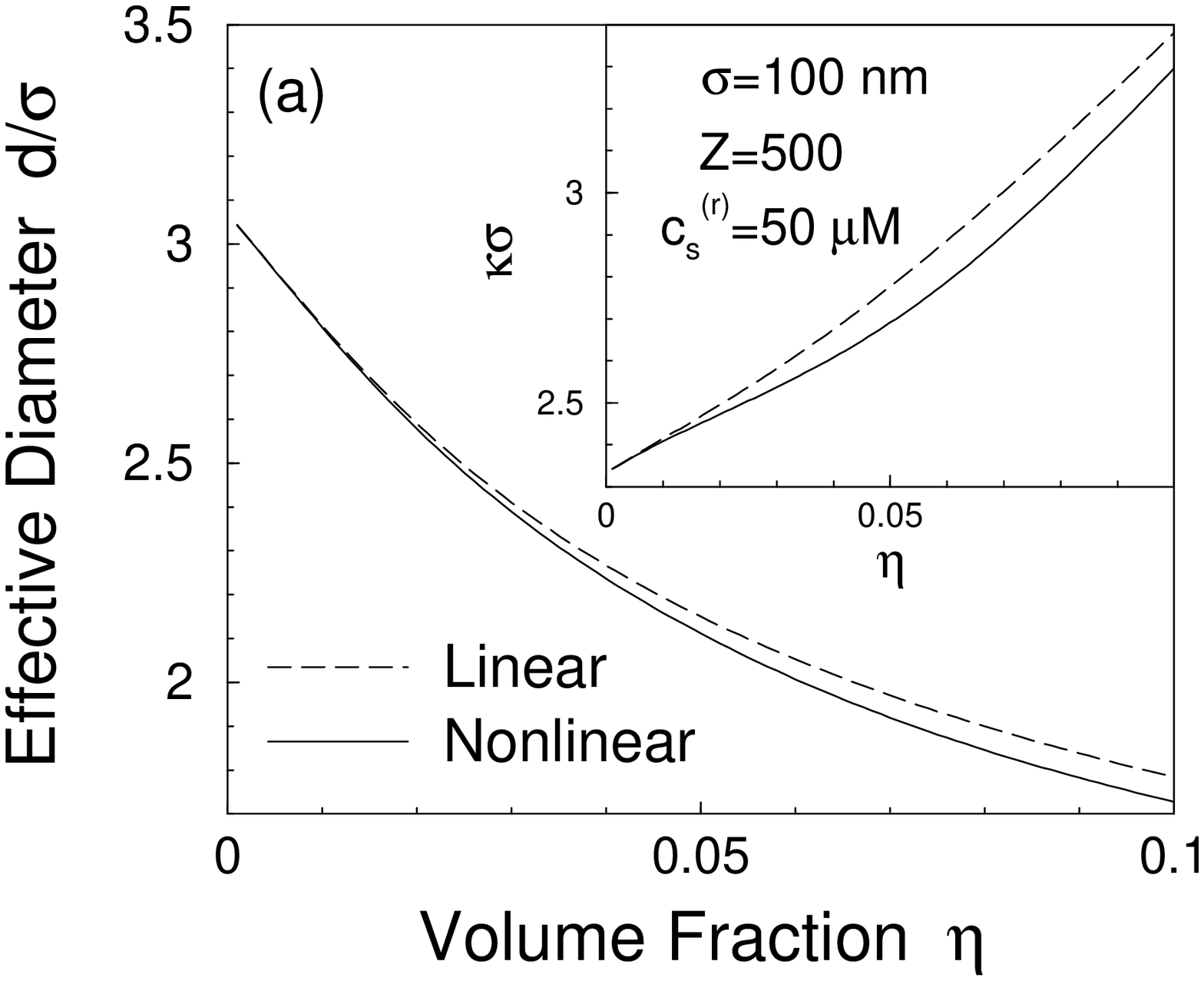} \\[2ex]
\includegraphics[width=0.75\columnwidth]{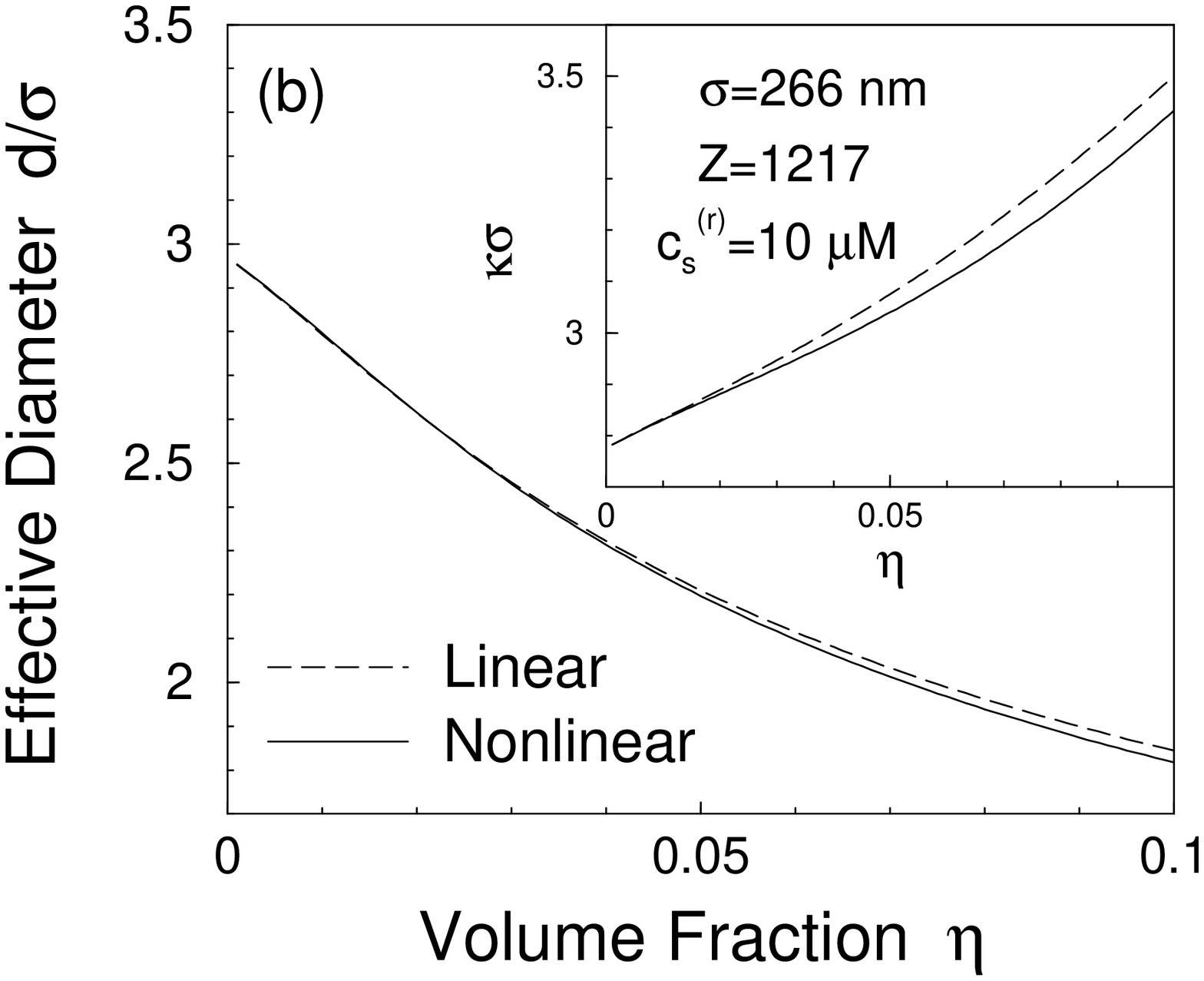}
\end{figure}

\begin{figure}
\includegraphics[width=0.75\columnwidth]{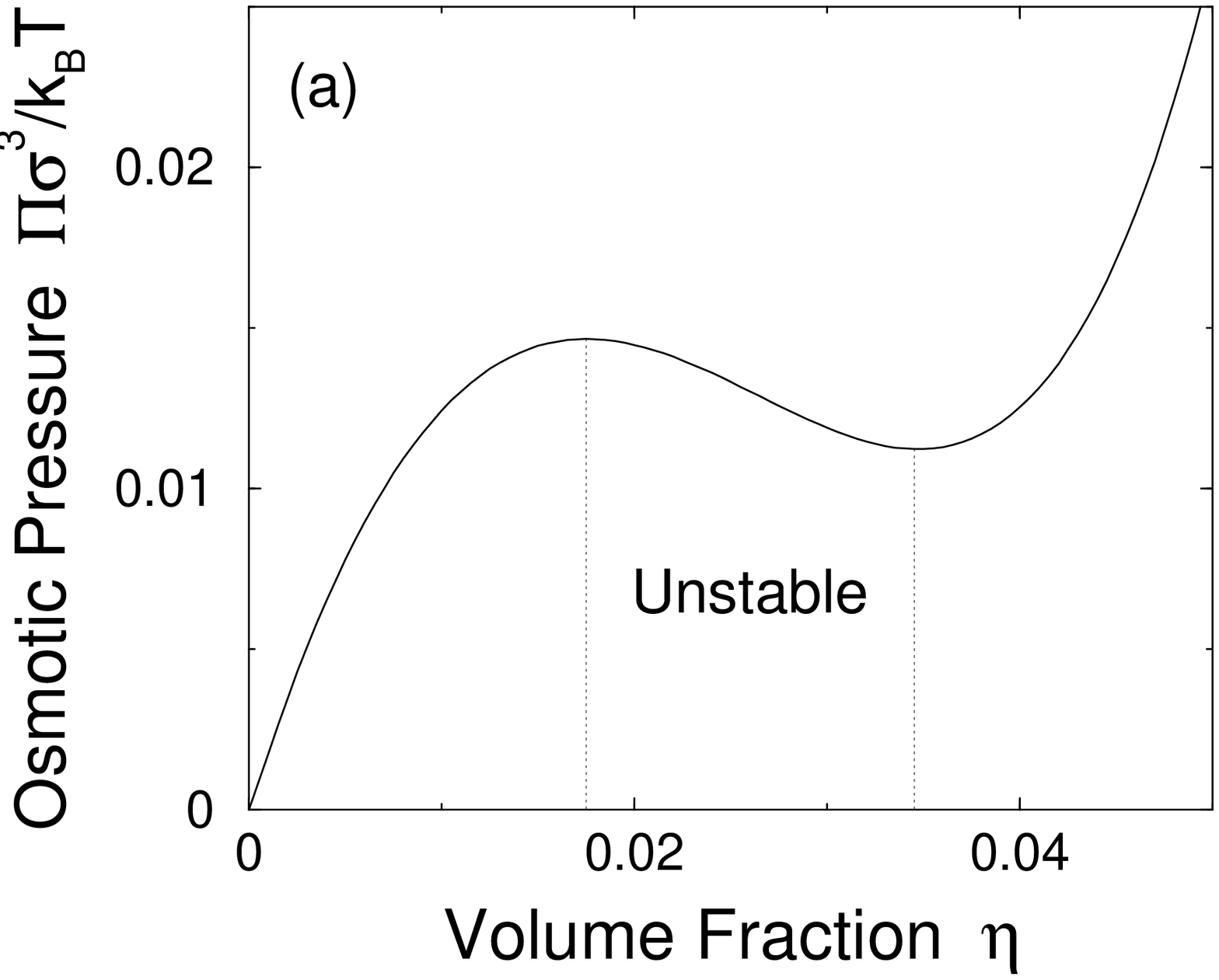} \\[2ex]
\includegraphics[width=0.75\columnwidth]{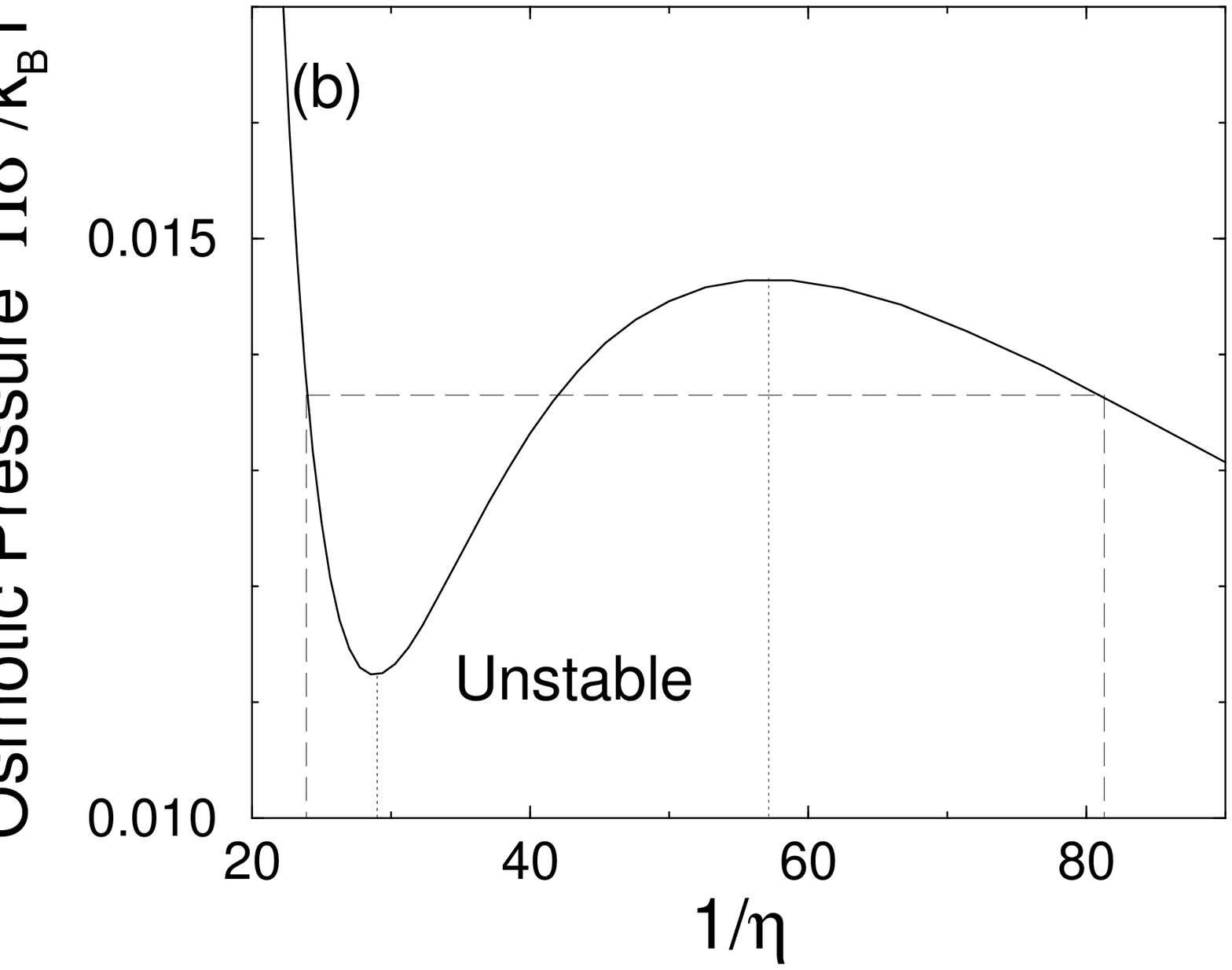} 
\end{figure}

\begin{figure}
\includegraphics[width=0.75\columnwidth]{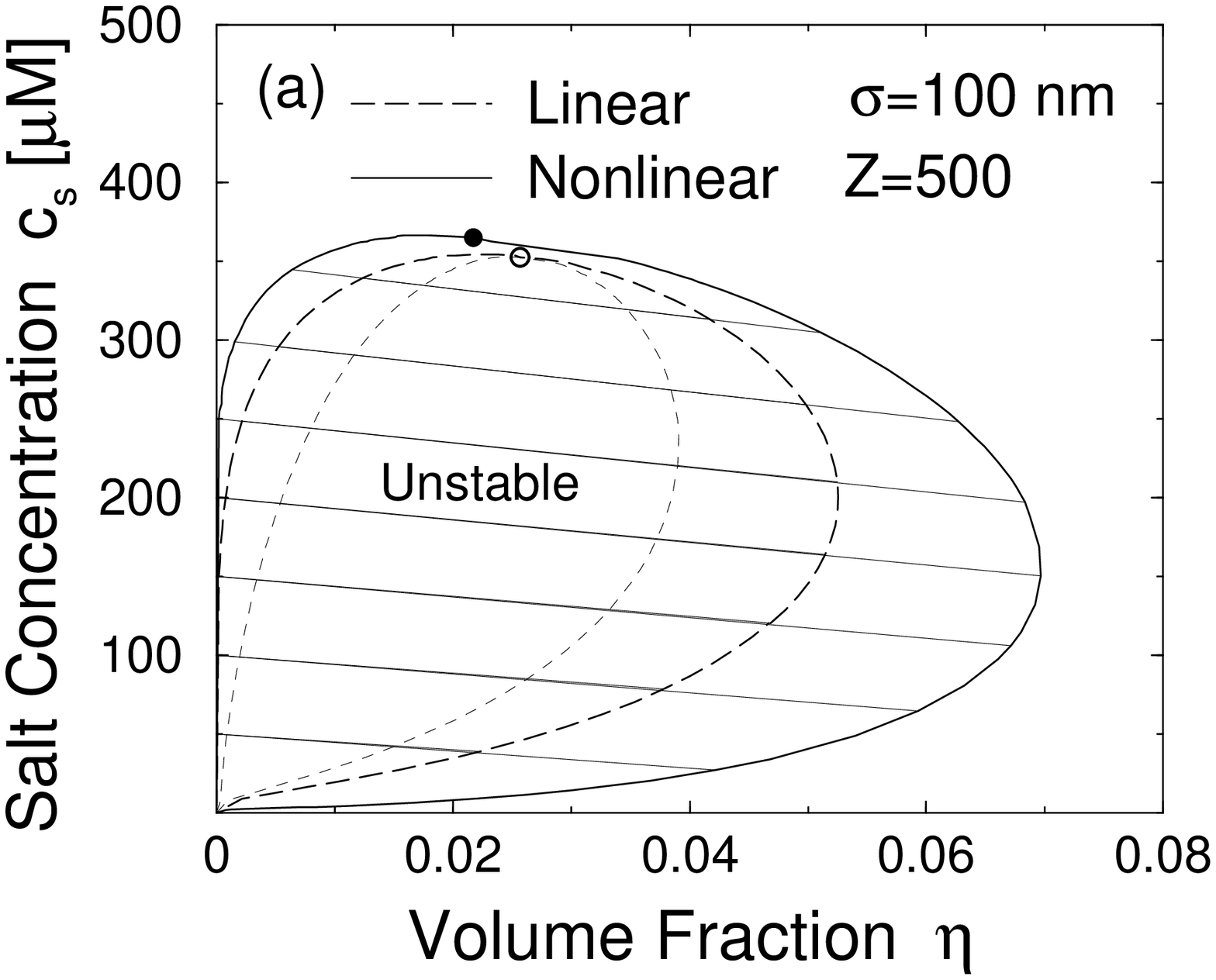} \\[2ex]
\includegraphics[width=0.75\columnwidth]{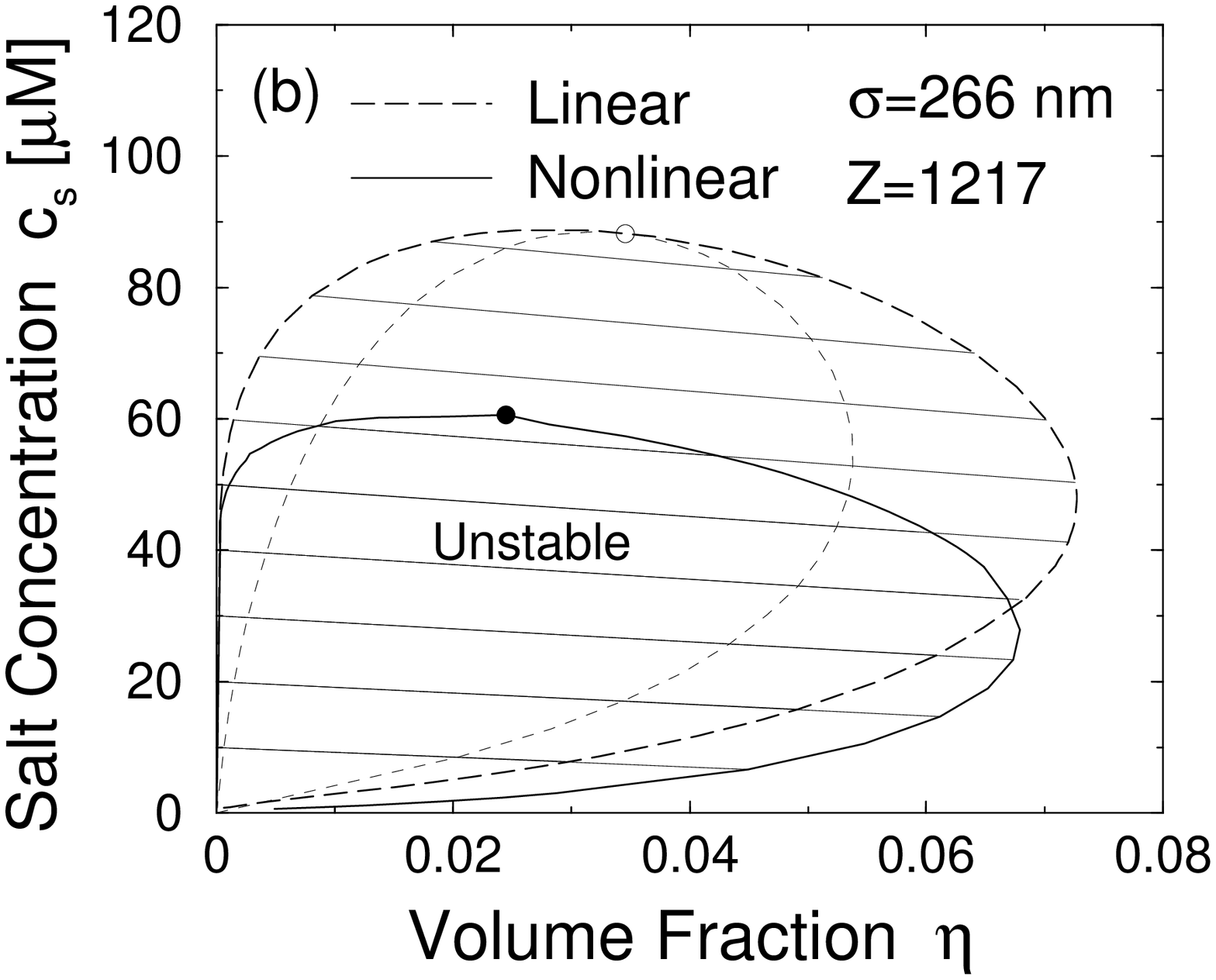} 
\end{figure}

\end{document}